\documentclass[trackchanges, twocolumn]{aastex701}

\graphicspath{ {figures/} {/}}

\begin{document}

\title{Stellar Flares Rather than High $T_{int}$ As Possible Cause for the Missing Methane in Young Exoplanets: The Case Study of V1298~Tau\,b}




\author[orcid=0000-0001-9897-9680,sname='Adams']{Danica Adams}
\affiliation{Department of Earth, Planetary, and Space Sciences, The University of California, Los Angeles, 595 Charles E. Young Drive East, Los Angeles, CA 90095, USA}
\email[show]{danica.adams@ucla.edu}  

\author[orcid=0000-0001-6809-3520, sname='Roy']{Pierre-Alexis Roy} 
\affiliation{Department of Earth, Planetary, and Space Sciences, The University of California, Los Angeles, 595 Charles E. Young Drive East, Los Angeles, CA 90095, USA}
\email[]{paroy@ucla.edu}  

\author[orcid=0000-0002-0298-8089,sname=Schlichting]{Hilke E. Schlichting}
\affiliation{Department of Earth, Planetary, and Space Sciences, The University of California, Los Angeles, 595 Charles E. Young Drive East, Los Angeles, CA 90095, USA}
\email[]{hilke@ucla.edu}  

\author[orcid=0000-0001-5578-1498,sname=Benneke]{Bj\"{o}rn Benneke}
\affiliation{Department of Earth, Planetary, and Space Sciences, The University of California, Los Angeles, 595 Charles E. Young Drive East, Los Angeles, CA 90095, USA}
\email[]{bbenneke@ucla.edu}

\begin{abstract}

Recent observations of V1298~Tau\,b, a warm sub-Neptune (R$=9.85 \pm 0.3\,R_\oplus$, M$=13.1 \pm 5.3, M_\oplus$, T$_\mathrm{eq} = 670\,$K) orbiting a young (10-30\,Myr) K star, show depleted methane and surprising abundances of CO$_2$ in its atmosphere. This makes the planet an interesting target to study the “missing methane problem”, where HST, Spitzer, and JWST have revealed many sub-Jovian planets with effective temperatures cooler than 800\,K to be depleted in methane compared to equilibrium chemistry predictions. On V1298~Tau\,b, an intrinsic temperature (T$_\mathrm{int}$) of 500\,K is required to explain the methane depletion if strong mixing from a warm interior is the reason for the missing methane, but this T$_\mathrm{int}$ is too large for a planet of its age and size according to standard interior structure evolution theories. Here, we show that stellar flares can result in methane depletion in a warm planet's atmosphere with nominal  T$_\mathrm{int} \sim 100$\,K. During a stellar flare, fast methane photolysis rates cannot be balanced by diffusion from below into the upper atmospheric layers probed by observations. Young active stars flare frequently, and we show that the timescale of diffusion for the methane to recover from a flare-induced depletion can be slow compared to the flare frequency. Over time, the atmosphere reaches a new steady state set by the average flare frequency and energy. This new steady state shows depleted methane in the 2\,mbar region, and abundant CO, CO${_2}$, and HCN, which results in synthetic transmission spectra that reproduce the main features of the recent JWST observations of V1298~Tau\,b.

\end{abstract}


\section{Introduction} 

To date, atmospheric detections in transmission and emission are easiest for large exoplanets, closely orbiting their stars, with hydrogen-rich envelopes. While these worlds are very different from planets in our own solar system, they act as valuable case studies to expand our understanding of atmospheric chemistry and its implications for planetary interior structures and their evolution and formation history. One might expect the high temperatures of warm atmospheres to lead to fast chemical reaction and to overall equilibrium chemistry, but evidence of disequilibrium chemistry has been discovered on many exoplanets \citep[e.g.][]{moses_disequilibrium_2011,roudier2021,baxter2021,tsai2023so2}.

The Hubble Space Telescope (HST) revealed a large mystery in the exploration of exoplanet atmospheres: the “missing methane problem”. From equilibrium chemistry, objects with H$_2$-rich atmospheres at temperatures cooler than 800 K are expected to have large abundances of methane \citep[][]{moses_chemical_2013}, and in the solar system all of the gas giants, albeit, cooler and further from their host star than recent exoplanet targets, have abundant methane. However, HST discovered that some of these objects lacked a methane feature that was expected from equilibrium chemistry models \citep[][]{stevenson_possible_2010,wakeford2018,chachan2019, benneke_sub-neptune_2019, spake2018, carone2021, baxter2021, thao2023}. Recently, JWST identified methane on a handful of exoplanets, including on sub-Neptunes whose HST's transit spectra had been interpreted as showing H$_2$O absorption instead of methane absorption \citep[][]{madhusudhan_carbon-bearing_2023, wogan_jwst_2024, beatty_sulfur_2024, benneke_jwst_2024, holmberg_possible_2024}. However, recent JWST measurements also support the missing methane problem, especially for planets in the 500-800 K temperature range, such as V1298~Tau\,b \citep[][]{barat_metal-poor_2025} and WASP-107b \citep[][]{sing_warm_2024, welbanks_high_2024}. To date, methane has only been detected on a few Neptune-sized worlds including sub-Neptunes K2-18b and TOI-270d \citep[][]{madhusudhan_carbon-bearing_2023, benneke_jwst_2024, holmberg_possible_2024}; warm Neptunes WASP-80b and GJ 3470b \citep[][]{bell_methane_2023,beatty_sulfur_2024}, and LP 791-18c \citep[][]{roy_diversity_2025}.  

One explanation for the missing methane is strong vertical mixing from a deeper, warmer region \citep[][]{stevenson_possible_2010, knutson2011, lanotte2014}. Vertical mixing is uncertain but may be strong at warm gas giants. GCMs predict the eddy diffusion coefficient ($K_{zz}$) at hot Jupiters to be $\mathrm{10^8-10^{12} cm^2/s}$ \citep[][]{moses2011, parmentier2013}, which is much larger than that at the surface of Earth ($\mathrm{10^5 cm^2/s}$; \citep[][]{liu1984}) and Venus ($\mathrm{3 \times 10^4 cm^2/s}$; \citep[][]{woo1981}). $K_{zz}$ likely decreases with depth \citep[][]{ackerman2001} and compared to hot Jupiters eddy diffusion at sub-Neptunes may be more modest with $K_{zz}$ ranging from $10^4$ near 100 bar to $10^8$ near 0.1 mbar \citep[][]{liu2026kzz}. The $K_{zz}$ may also be highly variable over timescales of ~30 days \citep[][]{liu2026kzz}. If strong vertical mixing from a deeper interior is why methane is frequently not detected, the deep interior must be depleted in methane and therefore must be very warm, so that equilibrium chemistry prefers the production of CO instead of CH$_4$. To deplete methane in the upper atmosphere (where observations occur near 1 mbar), the deep quench region must be warm enough to be CO-dominated \citep[][]{fortney2020}. Such warm interiors require large luminosities commonly parameterized as $\mathrm{T_{int}}$, which describes the contribution of formation to the luminosity of a planet such that $\sigma T_{int}^4 + \sigma T_{eq}^4 = F_{tot}$ where $\sigma T_{eq}^4$ describes the re-radiation of the stellar energy absorbed by the atmosphere and $F_{tot}$ is the total flux emitted by the planet \citep[e.g.][]{guillot2002}. Invoking a large Tint such as 500 K at V1298~Tau\,b \citep[][]{barat_metal-poor_2025} is often required to have a deep region inside the planet warm enough to explain non-detections of methane with equilibrium chemistry. 

However, formation and evolution models predict much cooler interior temperatures ($\sim$100-200~K), and recent works find that several targets require a Tint too large to be consistent with standard evolution models \citep[][]{barat_metal-poor_2025, yu2026}. To overcome this `Tint problem', dynamical interactions such as misalignment of planetary spin with its orbital angular momentum \citep[][]{millholland2020} or tidal heating from eccentricity dampening \citep[][]{agundez2014, morley2017, fortney2020} have been suggested as possible extra energy sources. This may be the explanation for some planets such as Wasp 107 b \citep[e.g.][]{batygin2025}, but it is less plausible for other planets. For example, a large spin obliquity ($>80^o$) would be needed to explain a $\mathrm{T_{int}}$ of 400 K at V1298~Tau\,b, a young planet with a relatively low eccentricity \citep[][]{barat_metal-poor_2025}.


In this work we investigate explanations for the missing methane that are consistent with cooler internal temperatures and hence agree with expectations from formation and evolution models. Since stellar activity is the strongest in young stars, we investigate photochemistry as the possible cause for the missing methane.

\subsection{The Case for Photochemisty}
Photochemistry is known to deplete methane and preferentially build up CO. Originally shown in brown dwarf atmospheres, the net reaction

\begin{equation}
    CO + 3H_2\leftrightarrow CH_4 + H_2 O
\end{equation}

is slow moving to the right due to the high binding energy of the double bond in CO, with the hydrogenation through methanol as the rate-limiting step \citep[][]{zahnle2014}. In warm temperatures, this disequilibrium chemistry favors the formation of CO and destruction of methane. But there are also kinetic inhibitions against oxidizing $\mathrm{CH_4}$ to CO at strongly irradiated planets \citep[][]{line2010}. Previous works have studied the disequilibrium photochemistry at sub-Neptunes such as K2-18b \citep[e.g.][]{jaziri2025} and GJ 1214b \citep[][]{kempton2012} but found the methane profile was similar to equilibrium predictions at and below 0.1 mbar. Observations probe near 1 mbar, and in these works photolysis in quiescent conditions is well balanced by vertical mixing in this region. However, stellar flares may increase the NUV radiation of a star and deplete methane to the layer observations probe, which is deeper than the layer that photochemistry typically influences methane abundances during quiescent stellar periods. 

Stellar flares are predicted to be most frequent at young stars \citep[e.g.][]{feinstein2020,feinstein2024}, and around K stars flare rates of 0.1-0.2 day$^{-1}$ are common \citep[e.g.][]{feinstein2024}. The effective temperature of a flare is highly variable and commonly near 10,000 K but can exceed 20,000 K \citep[][]{howard2020}. At Trappist-1, effective temperatures ranging from 3,000 to $>$10,000 K have been reported \citep[e.g.][]{howard_characterizing_2023,maas_lower-than-expected_2022}, but more massive stars are often associated with hotter flare effective temperatures \citep[][]{howard2020}. The flaring region of the star is also highly variable but may be ~10s of percent of the stellar surface \citep[][]{piaulet-ghorayeb_jwstniriss_2024}. Importantly, in the measured transit of V1298~Tau\,b (a warm sub-Neptune missing its methane in JWST measurements), the stellar flux increases abruptly in magnitude comparable to the planet’s transit suggesting strong stellar activity \citep[][]{barat_metal-poor_2025}.

\subsection{V1298~Tau\,b as the test case}
V1298~Tau\,b (R = 9.85 $\pm 0.35 R_\oplus$, Teq = 670 K) orbits a young (10-30 Myr; \citep[][]{david2019v1298,suarezmascareno2021, maggio2022, finociety2023} K star at an orbital period of 24.139 days \citep[][]{david2019v1298}. It is one of four planets in the system, and the inner three were found near a 2:3:6 mean-motion resonance \citep[][]{david2019v1298}. Since it is around a young star, RV is generally challenging due to interference from stellar activity \citep[][]{brems2019,tran2024, blunt2023}. \citet{livingston2026} finds a mass of $13.1 \pm 5.3 M_\oplus$ from TTV observations. From the atmospheric scale height, \citet[][]{barat_metal-poor_2025} retrieve a mass of 12 $\pm$ 1 $M_\oplus$, which is the mass we consider in our modeling work. Their estimates suggest a gas-to-core mass fraction of 0.1-8 percent with a core mass of 11-12 $M_\oplus$, suggesting the planet is a gas-dwarf sub-Neptune \citep[][]{barat_metal-poor_2025}. No atmospheric escape has been reported for this planet yet, but tentative detections were made at its neighbor V1298 Tau d \citep[][]{feinstein2021v1298, vissapragada2021, alam_jwst_2024}. Recent transit spectrum measurements with JWST NIRSpec/G395H suggest a metallicity of log Z = 0.6 +0.4,-0.6 $\times$ solar, detections of $\mathrm{CO_2}$ and $\mathrm{H_2O}$, and a relative lack of $\mathrm{CH_4}$ have been used to infer a C:O ratio of 0.23 \citep[][]{barat_metal-poor_2025}. Similarly, HST measurements suggest ~400 ppm water vapor in an extended and clear atmosphere, and missing methane was found \citep[][]{barat_metal-poor_2024}.

Since stellar activity is the strongest in young systems, we investigate the impact of photochemistry on their atmospheric composition. We couple photochemistry with radiative transfer to test whether the enhanced NUV flux from stellar flares can explain the missing methane at V1298~Tau\,b. In Section 2, we describe our methods for both photochemistry and radiative transfer. In Section 3, we describe photochemical model results during quiescent stellar conditions, and we explore the relevance of different eddy diffusivities ($K_{zz}$) and different metallicites and C:O ratios. In Section 4, we examine how the photochemistry changes during stellar flares, and we explore different flare durations, frequencies, and flaring temperatures. In Section 5, we discuss the implications of our V1298~Tau\,b results for other worlds with missing methane. We also discuss the lack of other photochemical products such as hazes and HCN in the measured spectrum of V1298~Tau\,b.

\section{Methods}
\begin{figure*}

\centering
\includegraphics[
    width=0.7\textwidth,
    trim=0 0.15cm 0 0.2cm,
    clip
]{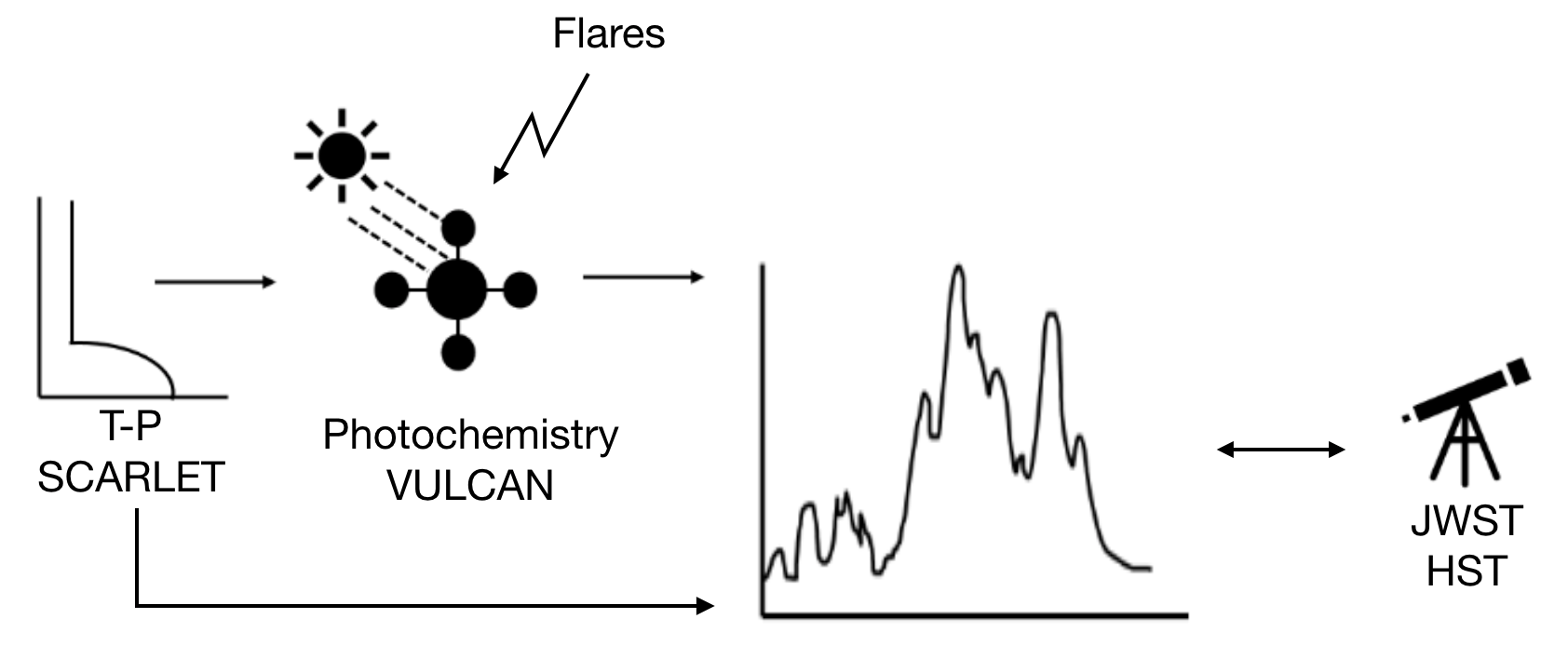}

\caption{Schematic of the experimental set up for this work. A temperature structure of V1298~Tau\,b in radiative-convective equilibrium is produced with SCARLET using a T$_\mathrm{int}$ of 100\,K and the chemistry from \citet{barat_metal-poor_2025}. This temperature structure is used as an input for VULCAN photochemical models in which the effect of flares on the photochemistry is also tested. The mixing ratios of the converged photochemistry models are then used along with the initial temperature structure to produce a modeled transmission spectrum with SCARLET, which we compare to the existing transmission spectroscopy observations from JWST and HST. }
\label{fig:fig01}
\end{figure*}

We self-consistently model the atmosphere of V1298~Tau\,b during quiescent and stellar flare conditions to make a direct comparison with the recent JWST observations (see Figure \ref{fig:fig01}). To do this, we couple a photochemistry code and radiative transfer code. The following Sections 2.1 and 2.2 describe each, and in Section 2.3 we describe their uses together.

\subsection{Photochemistry model: VULCAN}

In this work, we use VULCAN to model the chemistry of the upper atmosphere of exoplanets under different temperature conditions, bulk compositions, and stellar environments. VULCAN is a chemical kinetics code that lets the chemistry of an atmosphere evolve in time following a large network of chemical reactions along with their respective rates \citep{tsai_vulcan_2017, tsai_comparative_2021}. VULCAN also considers photochemistry i.e., the effect of incoming high-energy photons breaking apart some molecules by photolysis. Finally, transport is modeled using the eddy diffusion parameter ($K_{zz}$) and molecular diffusion. Other important inputs for VULCAN models include the temperature-pressure profile, the initial chemical composition, which is computed at equilibrium using FASTCHEM for given elemental ratios, and the planetary parameters (such as the semi-major axis and the gravity).

Throughout this work, we adopt VULCAN's CNOH network, which includes 800 reactions of carbon-, nitrogen-, oxygen-, and hydrogen-bearing species. We assume zero flux boundary conditions for all species except for hydrogen, which undergoes diffusion limited escape from the upper boundary. Every VULCAN model in this paper is generated using 120 pressure layers from $10^3$ to $10^{-6}$ bar.

\subsection{Radiative transfer: SCARLET}
We use the SCARLET framework \citep[e.g.][]{benneke_atmospheric_2012, benneke_how_2013, benneke_sub-neptune_2019, benneke_water_2019, pelletier_where_2021, piaulet2023, roy_water_2023, roy_diversity_2025} to model simulated transmission spectra of exoplanet atmospheres for different compositions and temperature structures. SCARLET parametrizes the molecular abundances, the temperature-pressure profile, as well as the cloud opacities, and then models the corresponding 1D atmospheric column consisting of 60 pressure layers in hydrostatic equilibrium before simulating the radiative transfer through the atmosphere and obtaining expected transmission spectra that can be compared to observations. The SCARLET framework has two modes that are leveraged in this work. 

In the forward modeling mode, we impose atmospheric conditions from which the 1D atmosphere model is produced, and observables are simulated. When comparing a forward model to transit observations, the 10 mbar radius of the planet is optimized by chi-square to match the observed transit depths. In particular, the forward model allows us to model the temperature structure of an atmosphere in radiative-convective equilibrium by solving the radiative transfer in the full non-gray scheme \citep{toon_rapid_1989}. In this setting, the energy budget (Bond albedo and heat redistribution factor), intrinsic temperature (T$_{\mathrm{int}}$), and the composition will dictate the converged temperature-pressure profile. 

In the retrieval mode, SCARLET fits atmospheric parameters to observed transit spectra within a Bayesian nested sampling framework \citep{skilling_nested_2004}. For each set of parameters tested by the sampling algorithm, SCARLET creates a 1D atmosphere model exactly as described in the previous paragraph. The planet radius at 10 mbar is optimized for each sample, and a gaussian likelihood is computed by comparing the simulated transit spectrum to the observations. The models are created at a resolving power of 20,000. This retrieval framework allows us to obtain the posterior probability distribution on atmospheric parameters (like molecular abundances, temperature, cloud deck pressures, etc.).

\subsection{Modelling V1298~Tau\,b using VULCAN and SCARLET}

We couple the VULCAN and SCARLET frameworks described above in order to perform a detailed study of V1298~Tau\,b, allowing us to couple the complex photochemical networks of VULCAN with the self-consistent radiative-convective temperature structures and simulated transmission spectra produced by SCARLET. We start by producing a SCARLET non-gray temperature profile in radiative-convective equilibrium for V1298~Tau\,b which serves as the input for our VULCAN photochemical models of the planet. We use the composition inferred from \citet{barat_metal-poor_2025} for the SCARLET model, i.e., a metallicity of 10$\times$solar and a solar carbon-to-oxygen ratio of 0.55. In terms of the energy budget, we assume a uniform planet (heat redistribution factor of 0.25), a Bond albedo of 0, and we use an intrinsic temperature of T$_\mathrm{int}$=100\,K which is consistent with formation and evolution models of V1298~Tau\,b \citet{barat_metal-poor_2025}. Since the atmospheric composition at chemical equilibrium depends on the temperature and pressure conditions, and since the temperature at a given pressure, in turn, depends on the atmospheric composition (via the opacity of the atmosphere), SCARLET's radiative transfer solver iterates between updating the temperature structure and the chemistry (recomputing the composition assuming chemical equilibrium at each step) until it finds its equilibrium temperature-pressure profile \citep{toon_rapid_1989}. We find that our SCARLET model is in broad agreement with the models presented in \citet{barat_metal-poor_2025}, both in terms of the temperature structure and the amplitude of the absorption features in the transmission spectrum.

We use VULCAN to model the upper atmosphere of V1298~Tau\,b including the effects of disequilibrium chemistry. We use the SCARLET temperature-pressure profile described above as an input for our VULCAN models. In order to model the irradiation environment of the planet, we parameterize the stellar spectrum during quiescent periods as a blackbody of 5000 K, which gives a near-solar luminosity of $~3\times10^{33}$ erg/s. In Section 3, we conduct sensitivity studies in quiescent stellar conditions by treating the following as free parameters: K$_\mathrm{zz}$, metallicity, and C:O. For these three parameters, we test cases with a range of $10^3-10^6$ cm$^2$/s, 1-10$\times$ solar, and 0.1-1 $\times$ solar, respectively. In Section 4, we simulate the effect of flares on the chemistry of the atmosphere of the planet by adding a stellar flare spectrum on top of the quiet stellar spectrum. The flare is described as a blackbody, and we treat the effective temperature and the effective emitting area of the flare (in fraction of the area of the stellar disk) as free parameters. In single-flare events, we consider effective temperatures of 6000, 12000, and 20000\,K, and we consider flaring areas from 1 to 10 percent. Motivated by the recent JWST observations, we ignore cases where the magnitude of the flare is significantly smaller or larger than the magnitude of the planet transit \citep{barat_metal-poor_2025}.

\subsection{Comparing photochemical models of V1298~Tau\,b to observations}

The recent HST and JWST transmission spectroscopy observations of V1298~Tau\,b (shown in Figure \ref{fig:bestfitmodel}) give us an important point of comparison for all the models tested in this work. For any converged VULCAN photochemical model, we can impose the temperature profile and the converged mixing ratios of the species in the atmosphere as inputs for a SCARLET forward model, allowing us to compare the expected transmission spectrum of the models to the existing observations. In order to make that comparison more informative, we must first correctly handle the presence of clouds in the atmosphere and the offsets between the different instruments that have observed V1298~Tau\,b thus far. 

We perform a SCARLET retrieval on the transmission spectrum published in \citet{barat_metal-poor_2025} in order to extract the cloud deck pressure as well as the instrument offsets that will be used when comparing VULCAN models to the observations. In our retrieval, we freely fit for the abundances of H$_2$O, CO, CO$_2$, CH$_4$, SO$_2$, NH$_3$ and HCN, all using log-uniform priors from 10$^{-10}$ to 1 in mixing ratio. We decide to perform our retrieval in this free chemistry mode in order to assess the presence of HCN which might appear as a consequence of the flares (Section \ref{sec:HCN}). We also fit for the pressure of a grey opacity deck (mimicking the effect of grey clouds) with a log-uniform prior from 1$\times10^{-6}$ to 10 bar, and we fit for an isothermal photospheric temperature using a uniform prior between 100 and 1200 K. Finally, we fit for offsets between the instruments used to assemble the spectrum: we anchor the transit depth on the JWST NIRSpec G395H/NRS1 detector, and thus we fit for an offset for the HST spectrum, and another offset for the JWST NIRSpec G395H/NRS2 spectrum. 

From our SCARLET retrieval, we find a pressure of 2 mbar for the grey cloud deck, and we find offsets of $417 \pm 5 $ ppm and $13 \pm 10 $ ppm for HST and JWST NIRSpec G395H/NRS2. The offset for the HST spectrum is also fitted in \citet{barat_metal-poor_2025}, and we find a fully consistent value. We create a new version of the transmission spectrum of V1298~Tau\,b to which we apply these instrument offsets, and which we can use for comparison with the VULCAN photochemical models. When using SCARLET to produce the simulated transit spectrum of a model with the VULCAN converged abundances, we adjust the 10 mbar radius of the planet to that new offset-adjusted spectrum, and we also set a grey opacity cloud deck at 2 mbar, the pressure empirically found in the retrieval. We then compute the chi-square between the models and the observations to assess whether they are a good fit to the data. Because of the difficulty of fitting the largely ill-behaved light curves of young stars such as V1298~Tau\,b (with large variations and flares), the transit spectrum of V1298~Tau\,b has underestimated uncertainties, leading to enlarged chi square values, even for best-fit models \citep{barat_metal-poor_2025}. Despite this, the large atmosphere scale height of this low-density planet and the high signal-to-noise ratio of the spectrum allow us to robustly detect the presence of absorbers in the atmosphere of the planet, even if the scatter in the measurements is larger than the uncertainties \citep[][]{barat_metal-poor_2025}. For that reason, we opt to use likelihood ratios computed from the rescaled $\chi^2$ of each model (rescaled so that the retrieval best-fit model has a reduced $\chi^2$ of 1) to compare the different scenarios we will produce in this work to one another and see how different effects improve or hurt the goodness of fit.

\section{Photochemistry in Quiescent Stellar Conditions}

\begin{figure*}
\gridline{
\includegraphics[
    width=0.99\textwidth,
    trim=0 0.5cm 0 0cm,
    clip
]{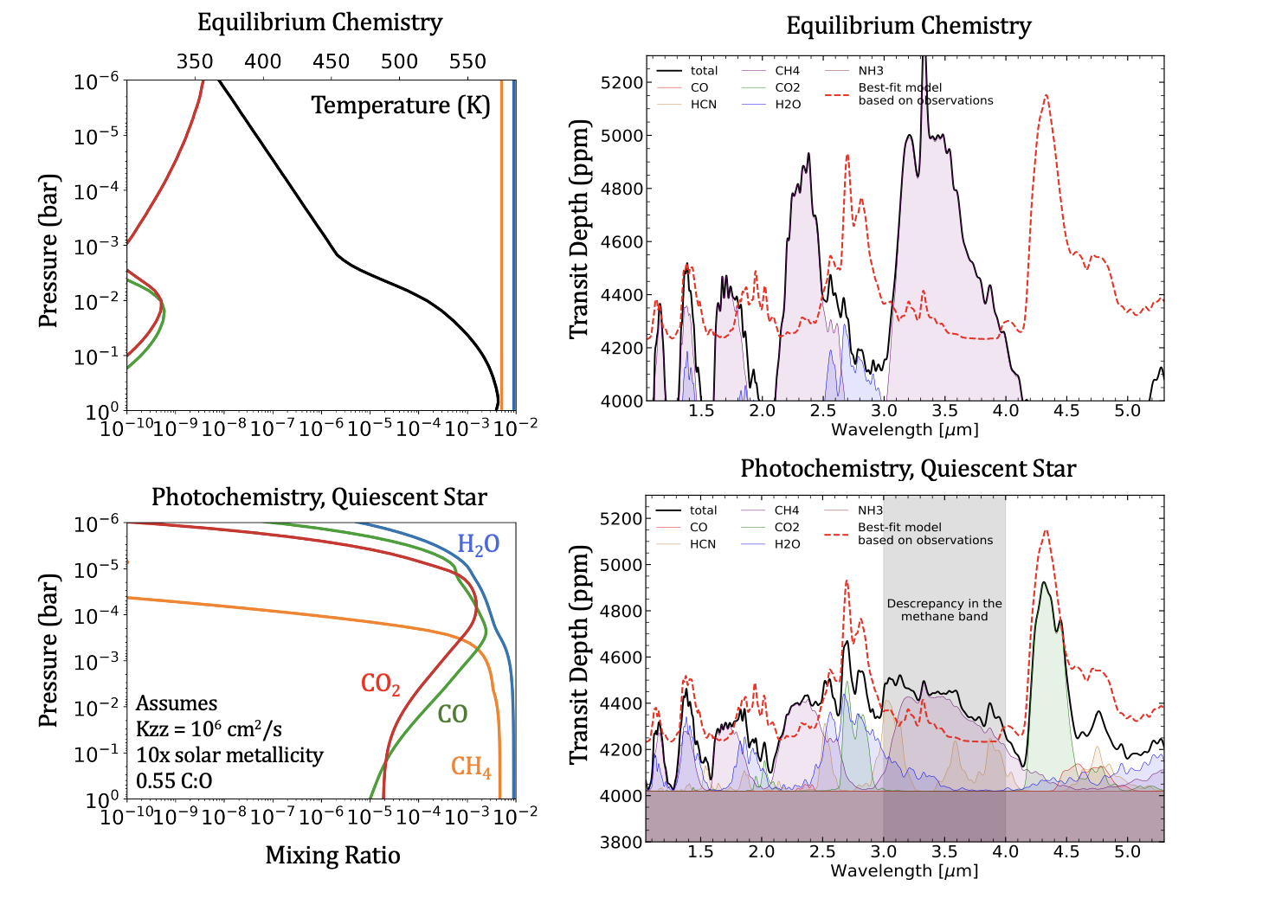}
}
\caption{Atmospheric mixing ratio profiles and their corresponding synthetic spectra for V1298~Tau\,b under standard equilibrium chemistry and photochemistry. (top left) Equilibrium chemistry (no photochemistry) profiles for $\mathrm{H_2O}$ (blue), $\mathrm{CH_4}$ (orange), $\mathrm{CO}$ (green), and $\mathrm{CO_2}$ (red) at a temperature-pressure profile with $T_{int}=100 K$. (top right) Corresponding synthetic spectrum for the model at chemical equilibrium shown on the left. The modeled transit spectrum is shown in black with the contributions from single molecules shown as the colored regions. The dashed red curve shows the best-fitting transit spectrum to the JWST and HST observations. (bottom left) Mixing ratio profiles from VULCAN (with photochemistry) for a quiescent star. This model assumes $K_{zz} = 10^6 cm^2/s$ with an atmosphere of 10 $\times$ solar metallicity and C:O ratio of 0.55 (i.e., solar). (bottom right) Same as in the top right, but with photochemistry. Both models over-predict the $\mathrm{CH_4}$ abundance near the observing layer (approximately 2 mbar). As a result both spectra show large $\mathrm{CH_4}$ absorption bands, mainly at 3.3\,$\mu$m,  which are inconsistent with the data.}

\label{fig:eqmix}
\end{figure*}

We start by first investigating the photochemistry at V1298~Tau\,b under quiescent stellar conditions to determine if the missing methane could be explained by photolysis.

Equilibrium chemistry predicts large, vertically-uniform abundances of methane throughout the atmosphere, which create large methane spectral features that are completely inconsistent with the recent JWST observations \citep[][Figure \ref{fig:eqmix}]{barat_metal-poor_2025}.  With photochemistry, photolysis (the primary sink of methane) is mainly balanced by diffusion from below (the primary source) in the observationally accessible part of the spectrum. Methane abundances decrease rapidly with height in the region above 1 mbar. As shown by our results in Figure \ref{fig:eqmix}, during quiescent stellar activity for an atmosphere with 10 $\times$ solar metallicity, a solar C:O ratio, and a modest $K_{zz}$ of $\mathrm{10^6 cm^2/s}$, photochemistry still overpredicts the spectral features of methane and underestimates those of $\mathrm{CO_2}$ and $\mathrm{CO}$. Since these results may be sensitive to $K_{zz}$, atmospheric metallicity, and C:O ratio, we vary these three parameters in the sub-sections below, to see if they alone could explain the observed depleted methane abundance.

Methane photolysis may form CH, $\mathrm{CH_3}$, and $\mathrm{CH_2}$. CH from methane photolysis is susceptible to oxidation by water vapor, first to $\mathrm{H_2CO}$, and subsequent photolysis results in the formation of CO. Oxidation of CO by OH is balanced by photolysis of $\mathrm{CO_2}$, and this chemistry causes both CO and $\mathrm{CO_2}$ to exceed expected equilibrium abundances (Figure \ref{fig:eqmix}). Meanwhile, $\mathrm{CH_3}$ may react to form ethane, as it commonly does at Titan and Pluto \citep[][]{krasnopolsky1999,wong2015,willacy2022}, or it may react with N (from $\mathrm{NH_3}$ photolysis) to form HCN either directly or first through $\mathrm{H_2CN}$. HCN is more stable to photolysis compared to $\mathrm{C_2H_6}$ \citep[][]{adams2022_titan,rimmer2019}, so at this planet (closely orbiting its host star), HCN becomes the favored reservoir and reaches mixing ratios $>10^{-4}$ (Figure \ref{fig:eqmix}). $\mathrm{CH_2}$ may react with $\mathrm{CH_3}$ to form $\mathrm{C_2H_4}$. 

The rate of methane photolysis becomes slower with depth due to photon absorption. Photochemistry in quiescent stellar conditions with moderate eddy diffusion (K$_{zz}$ = $10^6$ cm$^2$/s) predicts approximately $10^{-3}$ $\mathrm{CH_4}$ at 1 mbar (approximately the depth that observations probe). Therefore, this chemistry predicts a detectable methane feature as shown in Figure \ref{fig:eqmix} and does not explain the recent JWST measurements. Compared to the best-fit atmosphere model from the SCARLET free retrieval, we find the photochemical model in quiescent stellar conditions to be disfavored with a log-likelihood ratio of 397, effectively tripling the $\chi^2$ of the model. In the following subsections, we explore modifications to the chemistry described thus far to determine if the missing methane could be better explained for different assumptions of the atmospheric properties. In Section 3.1, we vary $K_{zz}$ as a free parameter, and in Section 3.2 we test photochemistry with different atmospheric metallicities and C:O ratios.


\subsection{Sensitivity of the observable Methane abundances to vertical Mixing ($K_{zz}$)}

\begin{figure*}
\gridline{
\includegraphics[
    width=0.9\textwidth,
    trim=0 0cm 0 0cm,
    clip
]{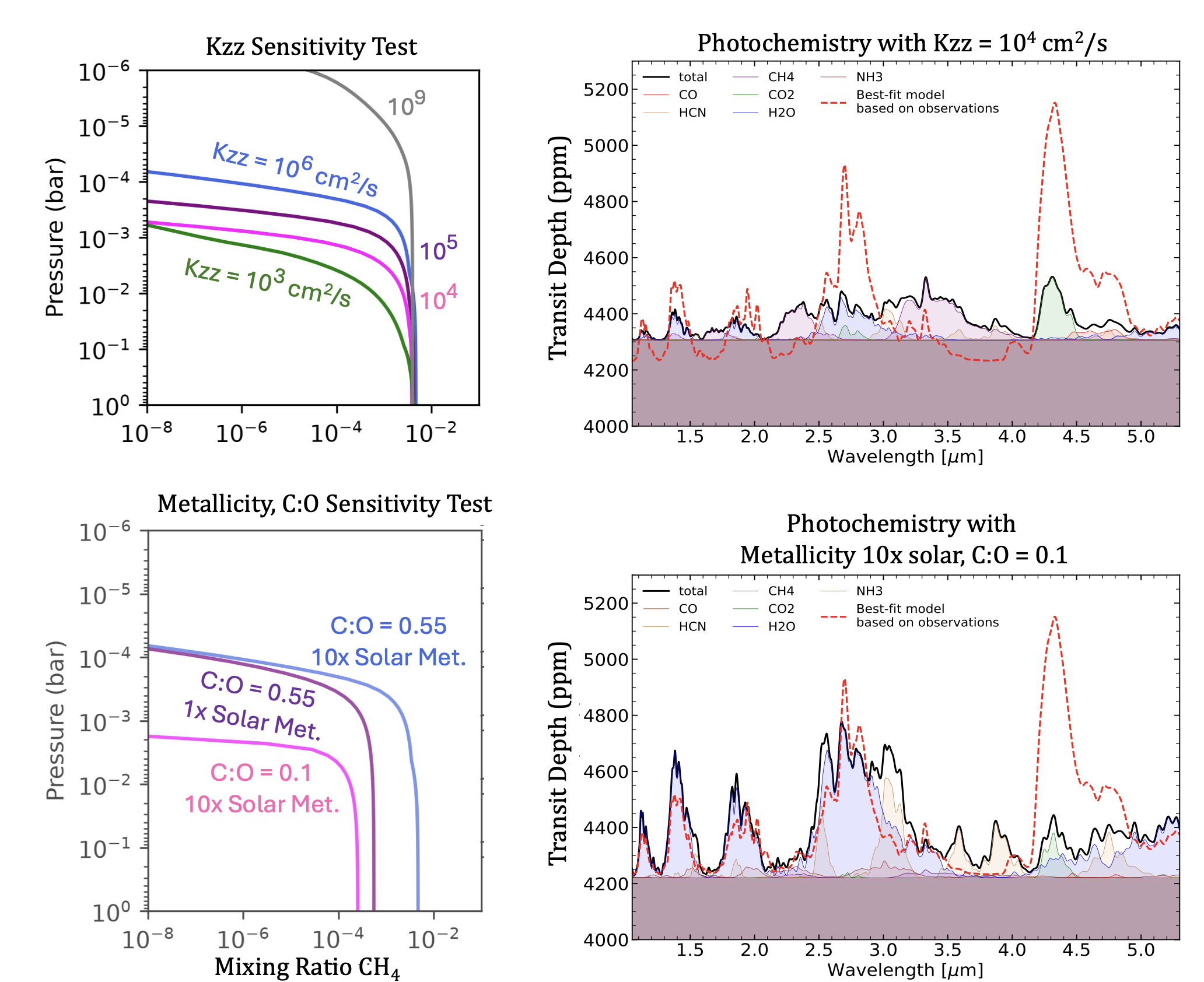}
}
\caption{Testing the effect of $K_{zz}$, metallicity, and C:O ratio on the missing methane. (top left) Mixing ratio profiles of methane for $K_{zz}$ of $10^6$ (blue), $10^5$ (purple), $10^4$ (magenta), and $10^3$ (green) $cm^2/s$. All cases assume 10 $\times$ solar metallicity and C:O of 0.55. (top right) Spectrum for the case using $K_{zz}=10^4 cm^2/s$. (bottom left) Mixing ratio profiles of methane for metallicity 10 $\times$ solar and C:O ratio = 0.55 (blue); metallicity 1 $\times$ solar and C:O ratio = 0.55 (purple); metallicity = 10 $\times$ solar and C:O ratio = 0.1 (magenta). (bottom right) Spectrum for case of 10 $\times$ solar metallicty and C:O ratio = 0.1. None of the cases investigated can explain the JWST data, so different choices of $K_{zz}$, metallicity, or C:O ratio together with photochemistry of a quiescent host star alone are not sufficient to explain the the observed depletion of methane.}
\label{fig:eddymettest}
\end{figure*}

Eddy diffusion (K$_{zz}$) parametrizes the efficiency of vertical mixing in a planet’s atmosphere, but remains poorly constrained. Values of $10^4-10^8 \mathrm{cm^2/s}$ have been assumed for sub-Neptunes of similar equilibrium temperature \citep[e.g.][]{yang_sub-neptunes_2026}, and recent GCM estimations of $K_{zz}$ profiles at K2-18b support this range \citep[][]{liu2026kzz}. As described earlier, the mixing ratio of methane is largely a balance of photolysis and vertical diffusion, so in atmospheres with longer mixing timescales, the atmospheric height where methane decreases rapidly to photolysis becomes deeper. However, we find this does not explain non-detections of methane because the opacity of other species in the upper atmosphere is also influenced. With slower mixing, CO and $\mathrm{CO_2}$ also deplete in the upper atmosphere, and observations probe deeper into the atmosphere. As a result, methane detections are still predicted by our chemistry models (Figure \ref{fig:eddymettest}).  

Previous works have shown that strong vertical mixing from a warm, deep region can explain the missing methane when invoking a high $\mathrm{T_{int}}$ \citep[][]{barat_metal-poor_2025,yu2026}. We are able to reproduce that result with a $\mathrm{T_{int}}\ge400 K$ and $\mathrm{K_{zz}}$ of $10^9 \mathrm{cm^2/s}$. In Figure \ref{fig:eddymettest}, we show that strong mixing ($\mathrm{K_{zz}}$ of $10^9 \mathrm{cm^2/s}$ with $\mathrm{T_{int}}$ of 100 K cannot explain the missing methane. The interior is not warm enough in this case, so equilibrium chemistry predicts $\mathrm{CH_4}$ in the material being lofted up by the strong eddy diffusion. The strong mixing lofts methane higher in the atmosphere since the rate mixing may now out-compete photolysis over a greater region in the atmosphere.

\subsection{Sensitivity of the observable Methane abundances to Metallicity and C:O Ratio}

The availability of oxidants and nitriles determines the fate of the CHx radicals that result from methane photolysis. At higher metallicity, the mixing ratio of all species increases, but the relative chemistry between them remains similar, so methane depletes from photolysis at a similar height between the 1x and 10x solar metallicity cases. However, the availability of oxidants and nitriles is modified by changes to the C:O ratio. In cases with a 0.1 C:O ratio, the relative amount of methane to water vapor is lower than in cases with 0.55 (solar) C:O. Oxidation of CH by water vapor leads to CO production, and oxidation of CO requires OH radicals which result from water photolysis. So, the formation of CO and $\mathrm{CO_2}$ are slowed in low C:O cases. As a result, less CO and $\mathrm{CO_2}$ are mixed to the upper atmosphere so photons are able to penetrate deeper in the atmosphere than in the modest C:O ratio cases. As a result, the depletion of methane in the upper atmosphere by photolysis falls to a lower height in the atmosphere ($\sim $1 mbar, Figure \ref{fig:eddymettest}). However, because photons are able to penetrate deeper, this methane remains detectable in the predicted transit spectrum. Therefore, decreasing the C:O ratio or metallicity cannot explain the missing methane. Another physical process must be responsible.

\section{Photochemistry and Atmospheric Response due to Stellar Flares}

We next examine stellar flares as a possible explanation for the missing methane, since we have shown that photochemistry in quiescent stellar periods for a range of atmospheric parameters could not explain the JWST measurements of V1298~Tau\,b.

Stellar flares may contribute to methane depletion since the NUV photon flux generally increases during a flare, and this is the spectral region that controls photolysis rates. The rate of photolysis is written as $J\times[CH_4]$, where J is the rate coefficient $\mathrm{s^{-1}}$ described as:

\begin{equation}
    J = \int \sigma_\lambda F_\lambda \gamma_\lambda d\lambda,
\end{equation}

where $\sigma_\lambda$ describes the photolysis cross section $cm^2$ at a given wavelength $\lambda$ (likelihood the molecule will interact with a photon), $F_\lambda$ is the photon flux at that wavelength, and $\gamma_\lambda$ is the quantum yield (describes the fraction of interactions that result in that reaction branch of dissociation). The lifetime of methane to photolysis, or equivalently the timescale of photolysis denoted $\tau_J$ hereafter, can be written as 1/J in units of $\mathrm{s}$.

\begin{figure*}
\gridline{
\includegraphics[
    width=0.9\textwidth,
    trim=0 0.7cm 0 0cm,
    clip
]{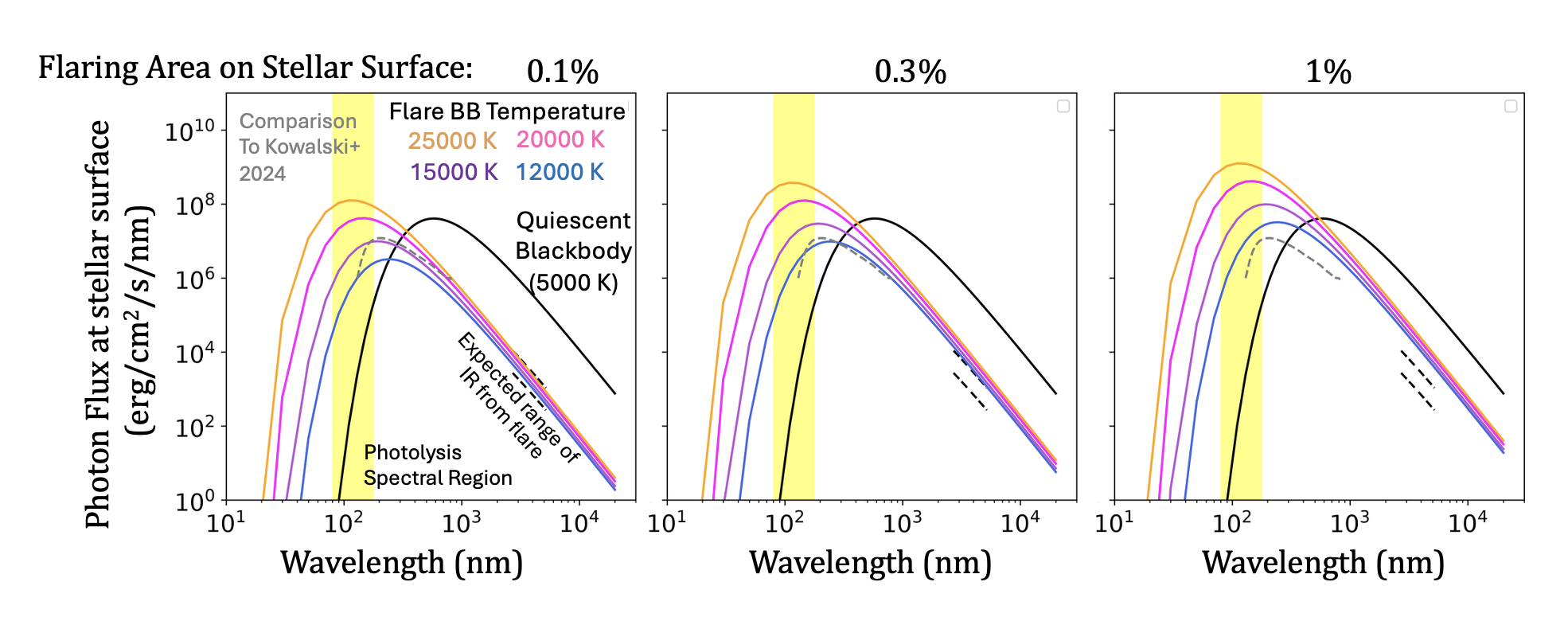}
}
\caption{Construction of the flare spectral energy distributions used in this work. We parametrize the stellar spectrum during a flare as the sum of the quiescent black body spectrum (5000 K emitted by the entire star) and the flaring spectrum, which we model with its own effective temperature and physical size (fraction of stellar surface area). Black body spectra of the quiescent star (5000 K, black) and the flare at effective temperatures of 12000 K (blue), 15000 K (purple), 20000 K (pink), and 25000 K (orange). The different panels describe different assumed areas of the flaring region on the stars surface increasing from 1 (left) to 3 (middle) and 10 percent (right). The yellow shaded region shows the spectral region relevant to photolysis. The dashed black lines show the expected contribution of the stellar flare to the IR flux according to the JWST measurement of V1298 Tau \citep[][]{barat_metal-poor_2025}. We compare (dashed grey line) to TESS measurements of M stars to demonstrate the energies of our flares agree in our 0.1\% area cases \citep[][]{kowalski2024megaflares}; in cases with larger flares, our model exceeds those measured energies. Measurements of K stars (hotter and more massive than M stars) are limited, and flares are stochastic by nature; so energetic flares may be plausible too. During a flare, the spectrum is modeled as the \textbf{sum} of the flare and the quiescent spectra shown here.
}
\label{fig:flaredef}
\end{figure*}

We parametrize the flare spectrum as a black body with its own temperature and relative surface area imposed over the star’s quiescent black body spectrum, similar to the parametrization in \citet{paudel2024}. As described in Section 1, the effective temperature of a flare and the ratio of the star’s surface that is flaring are both highly variable parameters for which we explore a range of plausible values. We consider effective temperatures for the flare of 12000, 15000, 20000 K, and 25000 K and we consider flaring areas of 0.1, 0.3 and 1 percent of the star's surface. These luminosities are of order $10^{32}-10^{34} \mathrm{erg/s}$, which range from an order of magnitude less to up to comparable to the stellar luminosity in quiescent periods ($\mathrm{3\times 10^{33} erg/s}$). This range agrees with the luminosities measured by TESS and Kepler during flares \citep[][]{gunther_stellar_2020,doyle2020,maehara2015}. The photolysis cross sections for our relevant species generally peak shortwave of Ly-A, making the hotter effective temperatures resulting in faster photolysis. The spectra of these modeled flares are shown in Figure \ref{fig:flaredef}. Stellar flares are known to increase the NUV flux. Different stellar atmospheric contributions cause shorter spectral regions to become best described by increasing effective black body temperatures, sourced by transitions and thermal responses in the upper atmosphere of the star \citep[][]{kowalski2024review}. TESS measurements of M dwarf CR Dra show the flux near 200 nm during a flare is much greater than that of a 10,000 K black body \citep[][]{kowalski2024megaflares} FUV Galex measuerments show the NUV flare from early M stars is also underestimated by a 9,000 K black body. \citet{berger2024} suggests the peaks of black body spectral energy distributions in the FUV are near 10,000-30,000 K. These works motivate our selected range of effective temperatures for the flaring region. We motivate the area of the flaring region (fraction of the star's total surface area) by comparing to the infrared change in the spectrum during the JWST measurement of V1298~Tau\,b's transit. An increase in IR approximately comparable to the size of the planet transit was observed \citep[][]{barat_metal-poor_2025}, and this constrains the IR contribution of the flare. In Figure \ref{fig:flaredef}, we show our modeled photon flux matches both the short wavelength  TESS measurements from \citet[][]{kowalski2024megaflares} and this IR change measured by JWST in \citet[][]{barat_metal-poor_2025}.

\begin{figure*}
\gridline{
\includegraphics[
    width=0.9\textwidth,
    trim=0 0.5cm 0 0cm,
    clip
]{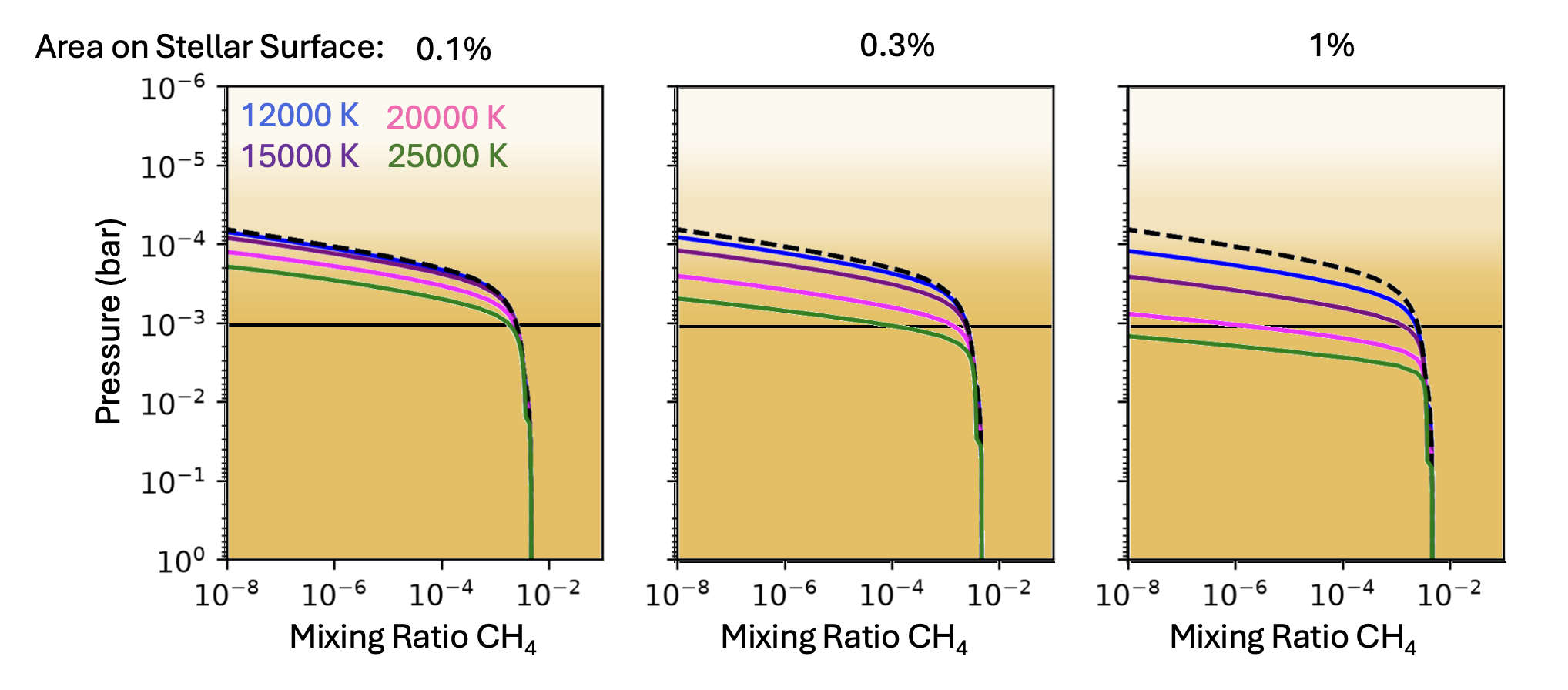}
}
\caption{Impact of different stellar flare temperatures and areas in single-flare-events on the atmospheric methane abundance profile. Panels left to right show different flare sizes (fraction of area on the stellar surface): 0.1, 0.3, and 1 \%. In each panel, the quiescent steady state result is shown in as a black dashed line. We initialize all flare cases to this steady state result, and then we change the stellar spectrum to the corresponding flare model and run VULCAN for $10^5$ seconds. Flare temperatures of 12000 K (blue), 15000 K (purple), 20000 K (magenta) and 25000 K (green) are shown in each panel. Observations probe close to 1 mbar, shown as horizontal black line. The background of each panel is shaded such that the greater opacity (below approximately 1 mbar) is where the atmosphere is opaque to observations. Above the black line at 1 mbar, the opacity of the shaded orange box decreases to depict the diminishing optical depth of the atmosphere. 
From these panels, we can see that single-flare cases that deplete methane to levels consistent with the JWST observations include: 0.1 \% fractional area and 25000 K ; 0.3 \% fractional area and 15000 K and 20000 K; 1 \% fractional area and 15000 K. Cases with larger flares over-deplete the methane compared to the data.}

\label{fig:oneshotflaremix}
\end{figure*}

\subsection{Atmospheric Response To Stellar Flares}

We find that stellar flares are effective at depleting methane in the layers most relevant to observations (approximately 2 mbar), and more energetic flares cause the most depletion. In Figure \ref{fig:oneshotflaremix}, the following four flares result in modest depletion, making them interesting cases to compare in more detail to the observation: (1) 25000 K of 0.1$\%$ stellar area; (2) 15000 K of 0.3$\%$ stellar area; (3) 20000 K of 0.3$\%$ stellar area; and (4) 15000 K of $1\%$ stellar area. Flares more energetic than these deplete methane potentially too strongly to match the data. We note here that Figure \ref{fig:oneshotflaremix} shows the methane abundance after a single flare event of $10^5$ seconds with constant luminosity. Single flare events are unlikely to last this long in reality. We simulate this duration to mimic the long-term effect of many sequential flare events and to predict which flare luminosities may be most relevant to model multiple shorter flares in detail over time in the next section. As we show in the Appendix with a toy model, the frequency of repeated flare events rather than each individual duration governs the steady state mixing ratio abundance near the observable atmosphere.


The following physics governs our result in Figure \ref{fig:oneshotflaremix}. For stellar flares to deplete the methane abundance at a given layer in the atmosphere, methane’s lifetime against photolysis ($\tau_J$, described in Section 4) must be shorter than its timescale for vertical diffusion ($\tau_{Kzz}$) and these timescales must be comparable or shorter than the duration of the flare ($\tau_F$):
\begin{equation}
    \tau_J \leq \tau_{Kzz} \leq \tau_F
\end{equation}

\begin{figure*}
\centering
\includegraphics[
    width=1.0\textwidth,
]{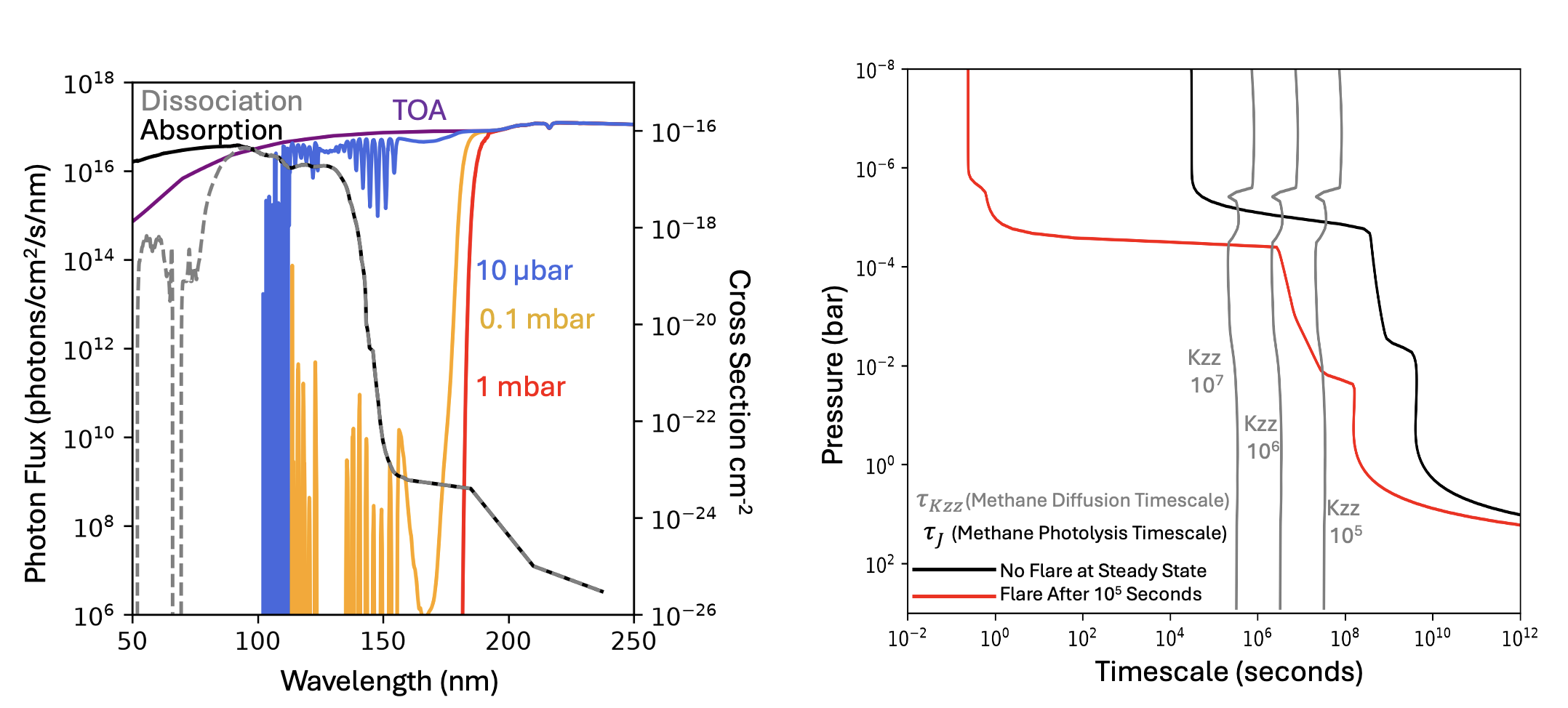}
\caption{Illuminating the key physics underlying the atmospheric mixing ratio profiles: the competition between photolysis and diffusion. Photolysis rates are determined by cross section and photon flux available at each level in the atmosphere after accounting for scattering and absorption (actinic photon flux). (left) Actinic photon flux at different levels of the atmosphere (top-of-atmosphere, TOA, in purple, 10 µbar in blue, 0.1 mbar in orange, and 1 mbar in red). These are compared with the $\mathrm{CH_4}$ cross sections (cm$^-2$) for absorption (black) and dissociation (dashed grey). Cross sections are shown on the y-axis on the right side of the figure. (right) Profiles of the $\mathrm{CH_4}$ photolysis reaction rate coefficients, J ($s^{-1}$) for methane dissociation from (black) quiescent no flare conditions to steady state and (red) after a $10^5$ second flare. The grey lines intersect the red and black ones where diffusion is competing with photolysis. Above the intersection, photolysis is faster, and below the intersection diffusion is faster.
}
\label{fig:tau_fig}
\end{figure*}

The timescale for diffusion is similar at all layers throughout the atmospheric profile in our simulations since we assume a constant $K_{zz}$ as previous photochemical studies have done for similar worlds \citep[e.g.][]{jaziri2025,kempton2012}. Realistically, $K_{zz}$ likely increases as $\frac{1}{\sqrt{P}}$ \citep[][]{ackerman2001,liu2026kzz}, so the timescale should be slower in the deep atmosphere and faster in the upper atmosphere. The timescale for photolysis is faster in the upper atmosphere where few photons have been absorbed by methane and other species such as CO, $\mathrm{CO_2}$, and $\mathrm{H_2O}$ (Figure \ref{fig:tau_fig}). The timescale increases sharply where the optical depth of most shortwave photons approaches unity, and at depth below this layer the timescale is long because only photons in the tail region of $\mathrm{CH_4}$ photolysis cross sections remain available to dissociate methane. 

Photolysis and diffusion are competing as the primary sinks and sources of methane when the timescales for diffusion and photolysis intersect, see Figure \ref{fig:tau_fig} (right panel). During stellar flares, the profiles for these timescales intersect deeper in the atmosphere than in quiescent periods. Above the point of intersection, i.e. higher in the atmosphere (Figure \ref{fig:tau_fig}), photolysis is much faster, which is why the methane mixing ratio profiles quickly deplete in the upper atmosphere (Figure \ref{fig:oneshotflaremix}). Below the point of intersection, there is a range of pressure levels where diffusion is faster but photolysis timescales are still relevant. This is where the atmospheric opacity is slowing the rate coefficient of photolysis; ie, photons have already been absorbed higher up, so photolysis is slower at depth. But, enough photons are penetrating that photolysis is an important sink of methane. There is a second `knee' in the timescale profile near 10 mbar where the atmosphere becomes more opaque to photons. In this case, mixing is much faster than photolysis (Figure \ref{fig:tau_fig}), and the methane profile is nearly constant (Figure \ref{fig:oneshotflaremix}) at a mixing ratio close to what equilibrium chemistry would predict. This methane is being vertically mixed up from a deep region where equilibrium chemistry is a good approximation, and the sink is slow in this deeper region.

\subsection{Multiple Flares Continuously Deplete Methane}

As mentioned above, young stars are known to flare frequently, so repeated flare events are likely a frequent occurrence in young exoplanet systems \citep[e.g.][]{feinstein2020}. We hypothesize that between flare events the methane abundance will remain suppressed due to long vertical mixing timescales into the upper regions recently depleted by a flare. 

To test this hypothesis, we couple a series of Vulcan simulations together into a timeseries that alternates between flares and quiescent periods. We initialize each simulation with the steady-state mixing ratio profile. We alternate between flare conditions ($10^5 $ s) and quiescent conditions ($4\times10^5$ s) for four flares of different energies. In the Appendix, we demonstrate with a simpler toy model (which reduces the long computational run-times) that the flare frequency influences our results but the absolute duration of the period does not.  We select the four flare energies based on the single-flare-event results shown in Figure \ref{fig:oneshotflaremix} and described in Section 4.1. Again, these cases include: (1) 12000 K of 0.1$\%$ stellar surface area; (2) 15000 K of 0.3 $\%$ area; (3) 20000 K of $0.3\%$ area; and (4) 15000 K of 1$\%$ area.

\begin{figure*}
\gridline{
\includegraphics[
    width=0.7\textwidth,
    trim=0 0.7cm 0 0cm,
    clip
]{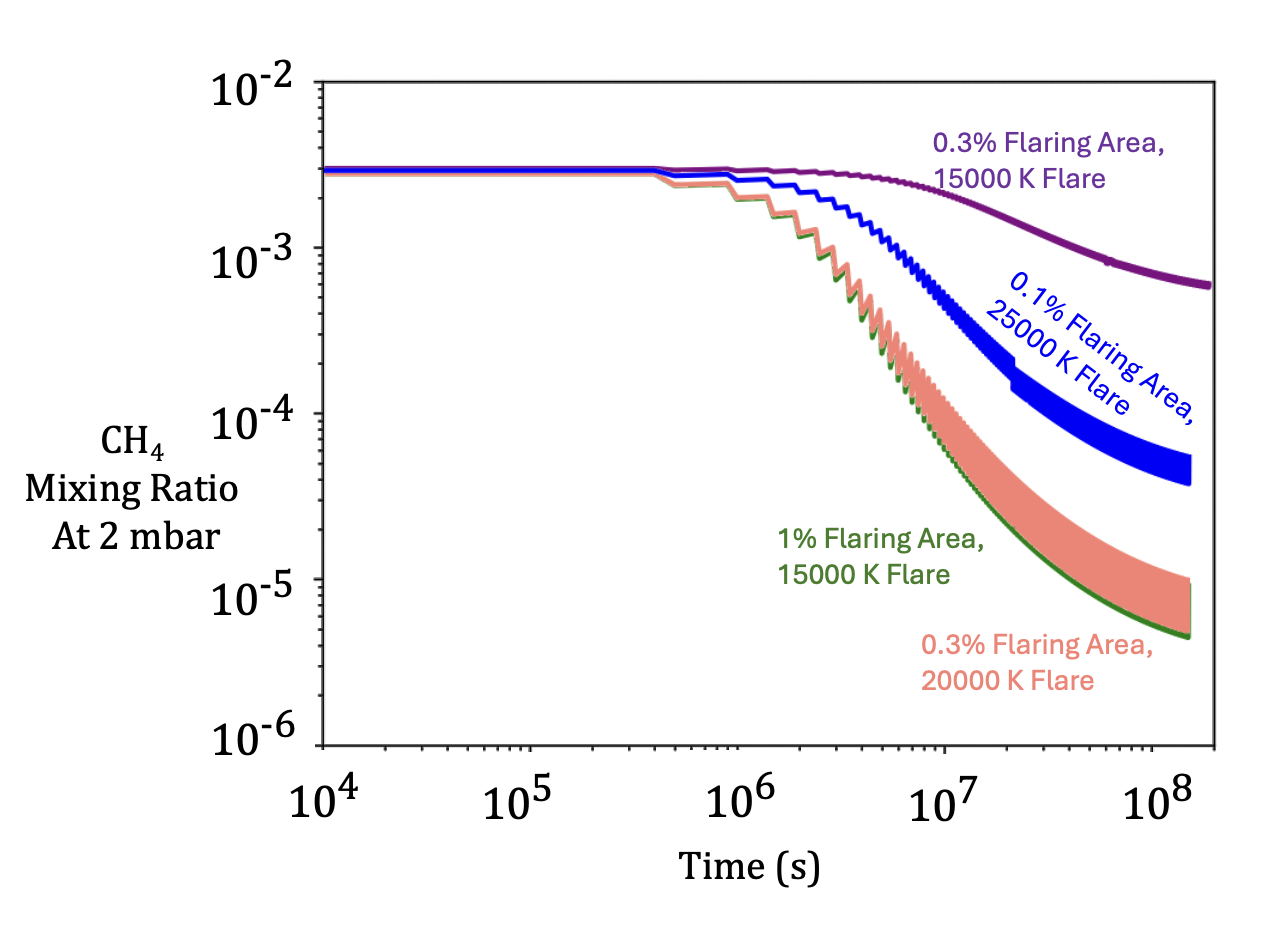}
}
\caption{ The photochemistry reaches a new steady state of methane abundance as the star changes between quiescent and flaring conditions over time. VULCAN time-series results of methane mixing ratio at 2 mbar for flares with 20 percent frequency ($10^5$ s flares separated by $4\times 10^4$ s quiescent periods). Four cases are shown: 0.3 percent of the stellar area at 15000 K, 0.1 percent of the area at 25000 K (blue), 0.3 percent of the stellar area at 20000 K (peach), and 1 percent of the area at 15000 K (green). Repeated stellar flares can explain the missing methane at V1298~Tau\,b, and the magnitude of depletion is sensitive to the frequency and energy of the flares.}
\label{fig:togglevulcan}
\end{figure*}

After the first simulated flare, methane depletes rapidly in the upper atmosphere ($P<0.1$ mbar), is unaffected in the lower atmosphere ($P>10$ mbar), and shows a modest decrease between these regions (which spans the region where observations probe). This pressure-dependence results from the column photon absorption above a given layer that influences the timescale of photolysis. During the quiescent period that follows, the methane in now depleted regions attempts to replenish by vertical mixing from deeper layers below. The diffusion timescale is relatively long, however, and the methane cannot fully recover before the next flare. As the flare and quiescent periods alternate over time, the planet receives a time-averaged incident photon flux and despite fast oscillations in a single flare/quiescent cycle, the atmosphere begins to converge towards a unique steady state to oscillate about (Figure \ref{fig:togglevulcan}). 

\begin{figure*}
\gridline{
\fig{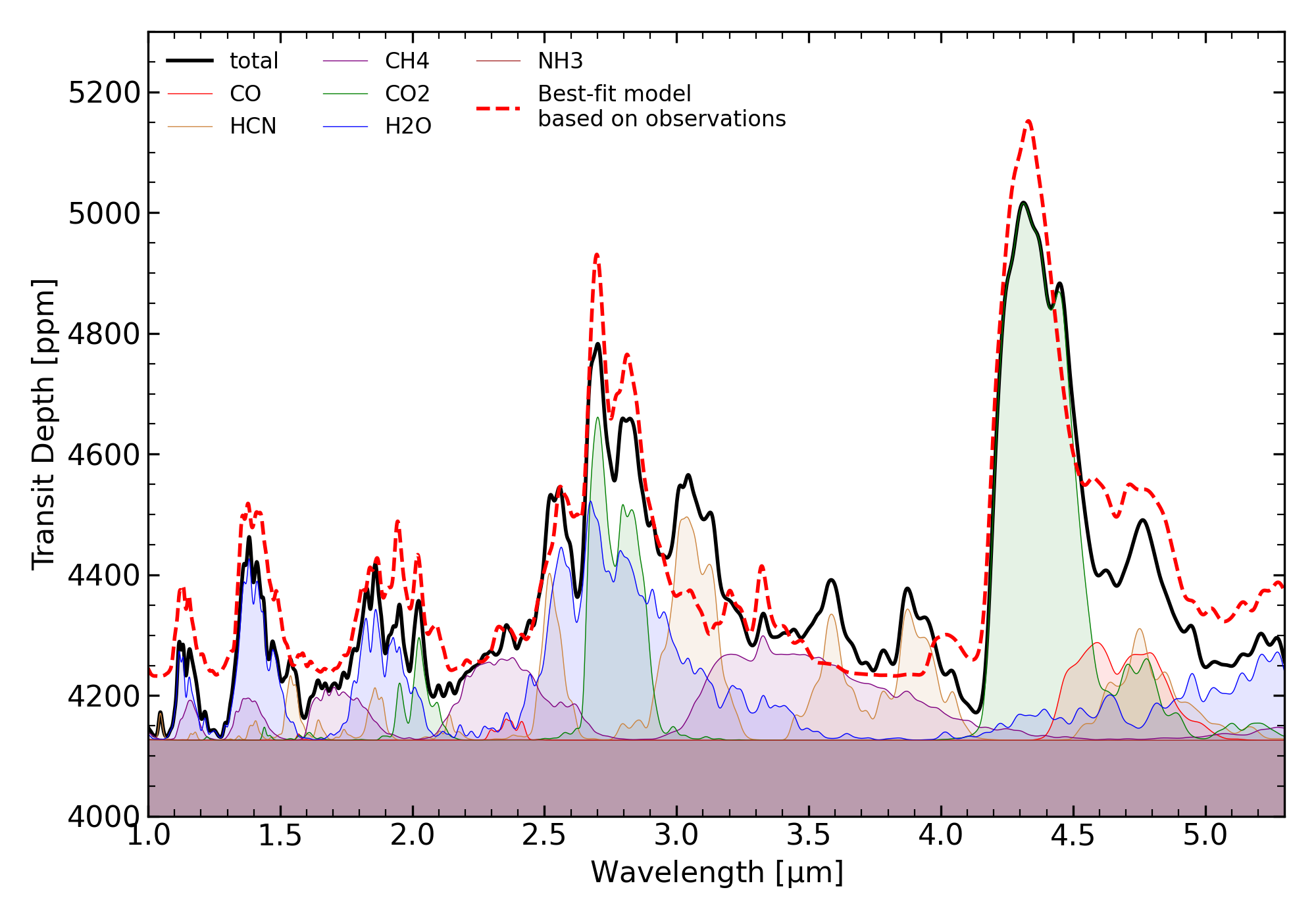}{0.99\textwidth}{}
}
\caption{Sequencing flare and quiescent conditions together can explain the missing methane at V1298~Tau\,b. In this case, VULCAN considers a flare frequency of 20 percent, effective flare temperature of 15000 K, and a flare size of 0.3 percent of the stellar surface. The peak methane absorption (purple) near 3.3 microns in the synthetic spectrum (black) matches the best-fit model on observations (dashed red) closer than the previous equilibrium and quiescent photochemical models shown previously. The $\mathrm{CO_2}$ absorption also increased in this model compared to the equilibrium chemistry and quiescent photochemistry models, making a better match near the 2.7 and 4.3 micron bands as well.}
\label{fig:bestfitspectrum}
\end{figure*}

Like the single-flare instances, the effective temperature of the flare and the ratio of stellar surface that is flaring both determine the incident photon flux during the flare and therefore control the photolysis rates. Figure \ref{fig:togglevulcan} shows a 20,000 K flare over 0.3 $\%$ of the stellar surface depletes methane just as efficiently as the 15,000 K flare over 1$\%$ of the surface. 

\subsection{Comparisons to VL 1298 Tau b}

As shown in Figure \ref{fig:bestfitspectrum}, we find that VULCAN is able to explain the missing methane on V1298~Tau\,b after a sequence of alternating flare and quiescent conditions. Our model where the flare has a temperature of 15000\,K, an area of 0.3\,\% of the star surface, and a 20\,\% flaring frequency is best at reproducing the current observations of V1298~Tau\,b. Compared to the best-fit model from the free retrieval, we find that this model is able to broadly reproduce the water absorption bands at 1.4 and 2.7\,$\mu$m, the CO$_2$ bands at 2.7 and 4.4\,$\mu$m, and even the depleted methane at 3.3\,$\mu$m (compared to chemical equilibrium expectations, or even that of photochemistry in quiescent conditions, Figure \ref{fig:eqmix}). Statistically, we find this stellar flare model to improve the $\chi^2$ of the quiescent photochemical model by a factor of 1.6, and to be favored with a likelihood ratio of $\ln \mathcal{L} = 200$. In terms of the CH$_4$ abundance, our free retrieval constrains a vertically constant mixing ratio of 5.79$\times10^{-5}$ above the 2 mbar cloud deck. At that mbar level, our forward VULCAN model finds a higher CH$_4$ abundance, around $5\times10^{-4}$, but this mixing ratio is quickly decreasing with lowering pressures above the cloud deck in that photochemically consistent model, explaining why both models show similar CH$_4$ features in the transmission spectrum.

In this multi flare model, VULCAN mildly underestimates $\mathrm{CO}$ and $\mathrm{CO_2}$. The abundances of $1.26\%$ and $0.26\%$ respectively constrained in our free retrieval are larger than the VULCAN mixing ratios of $0.274\%$ and $0.100\%$ respectively. Instead, the destroyed methane recombines with nitrogen in our VULCAN simulations, causing the HCN mixing ratio to climb to $0.116\%$, which exceeds our retrieved 2-$\sigma$ upper limit on the HCN abundance of $6.48\times10^{-6}$. In there lies perhaps the largest discrepancy between the observations, which do not seem to show the presence of HCN and our multi flare model, which does. Therefore the production and presence of HCN might be useful as an observational probe of atmospheric chemistry under flaring conditions. We however note that the N:H ratio of the planet is largely unconstrained and could play a role in the absence of HCN absorption from the observed spectrum (whereas our model assume the solar ratio). The water abundance in our model ($0.40\%$) underestimate the chemistry free-retrieval abundance ($0.87\%$), so a smaller C:O ratio and/or smaller N:H ratio may better fit the data. This possibility is consistent with the retrieved C:O ratio of $0.3$ in \citet[][]{barat_metal-poor_2025}.


\section{Discussion}

In this section, we discuss the implications of our results for other worlds known to have missing methane on a planet-by-planet basis before also discussing emerging observational trends and broader implications of our photochemistry results.

\subsection{Can Stellar Flares Explain the Missing Methane for other temperate Gas Giants?}

In addition to V1298~Tau\,b, several planets have low abundances or non-detections of methane despite equilibrium chemistry predicting large abundances. These include HD 189733b \citep[][]{fu_hydrogen_2024,zhang2025}, HIP 67522b \citep[][]{thao2024},  HAT-P-12b \citep[][]{crouzet2025}, HAT-P-18b \citep[][]{fu2022,fournier-tondreau_near-infrared_2024}, WASP-69b \citep[][]{schlawin2024}, GJ 3470 b \citep[][]{benneke2019gj3470}, WASP-107b \citep{kreidberg2018,dyrek2024}, WASP 39b \citep{ahrer_early_2023},TOI 421 b \citep{davenport2025}. The ages of their host stars show no clear pattern ranging from tens of millions of years to billions of years, so the frequency and strength of flares each world experiences likely differs. The planets are also of different masses and thermal structures, so the strength of vertical mixing in these planets’ atmospheres likely also spans a wide range. For example, \citet{adams2022} finds a $K_{zz}$ of approximately $10^{11} \mathrm{cm^2/s}$ at Kepler-7b, a warmer hot Jupiter, which is much larger than the $10^4-10^8$ range reported for sub-Neptunes \citep[e.g.][]{liu2026kzz}. Due to the uncertainty in K$_{zz}$, the depth of methane depletion during a flare event (where diffusion timescales are comparable to photolysis timescales) is uncertain in each atmosphere. In addition, the frequency of flares is much higher for younger stars than older stars, and during extended quiescent periods, the efficiency of replenishing methane from lower layers will vary according to each atmosphere’s $K_{zz}$ profile. GCM models of these different worlds could inform $K_{zz}$, and future photochemical studies of their chemistry should be done to test how frequent flares must be to explain methane depletion at these other worlds.

It is plausible the cause of the missing methane needs to be treated on a case-by-case basis. The original hypothesis of quenching only required an unphysically large Tint for some planets, whereas other worlds required a $T_{int}$ that was consistent with planet formation theories \citep[][]{yu2026}. In the following text, we discuss the implications of recent measurements at select worlds on a case-by-case basis. 

\subsubsection{GJ 3470 b}
GJ 3470 b is an interesting case where photochemistry from flares should be investigated. JWST measurements detected methane (depleted but present), and the detections of $\mathrm{H_2O}$, $\mathrm{CO_2}$, $\mathrm{SO_2}$, and $\mathrm{CH_4}$ suggest a subsolar C:O ratio \citep[][]{beatty2024}. This means oxidation pathways to CO and $\mathrm{CO_2}$ may be faster than forming larger hydrocarbons (haze). Importantly, earlier observations of GJ 3470 b resulted in several non-detections of methane \citep[][]{crossfield2013,biddle2014,benneke2019gj3470}, and disequilibrium chemistry, aerosols, and metallicity were all cited as possible explanations. The time variability of the methane in combination with its depletion may suggest the methane was destroyed by flares in the earlier measurements, and this also may suggest that the later JWST measurement observed the methane recovering post-flare. The age of its host star remains uncertain, reported as 0.3 to 3 Gyr \citep{bonfils_hot_2012,biddle_warm_2014}, but several episodes of flaring activity have been detected \citep{bourrier2021}.   

\subsubsection{HD 189733b}
There is also potential evidence for variability in methane abundance at HD 189733b. HD 189733 b is a warm target ($T_{eq}$ of 1209 K), so equilibrium chemistry already predicts low methane abundances. However, the detected abundance appears to be depleted, and a deep hot quench point is needed to explain the depletion with mixing and requires a  potentially unphysically high $T_{int}$ of 900 K \citep{yu2026}. The retrieved methane signal in ground-based NIR measurements of the dayside emission spectrum \citep[][]{swain2009} was larger than that of HST measurements of the terminator transmission spectrum \citep[][]{swain_presence_2008}. Both detections of methane co-occurred with detections of water vapor, and the former dataset also showed detections of $\mathrm{CO_2}$. \citet{swain2014} and \citet{gibson2011} highlight the importance of different reduction methods on the retrieved methane abundance, and the latter work suggested no methane was detected in the HST data. So it is unclear whether this variability is physical or a modeling artifact. If real, the larger abundance appearing on the day-side (warmer than the terminator) is chemically surprising and may suggest disequilibrium chemistry. In that case, potentially the day-side measurements observed the methane recovering post-flare, whereas the HST data could have sampled the planet at a time after a more recent flare. Improving the outlook for the case of flares at this world, flaring variability was measured to result in variable atmospheric escape, albeit at a different time period than either $\mathrm{CH_4}$ detection was made \citep[][]{lecavelier2012}.

\subsubsection{HIP 67522 b}
HIP 67522 b is another warm gas giant (Teq ~ 1176 K) that would require a high $\mathrm{T_{int}}$ (750 K) to explain the lack of methane \citep[][]{yu2026} measured by JWST \citep[][]{thao2024}. This planet orbits a young, active star of 17 Myr \citep{rizzuto2020}. HST/COS and XMM-Newton observations revealed strong variability in flaring, a hot corona, and a high XUV flux \citep[][]{maggio2024}. It has been suggested from TESS photometry that the flaring rate may be abnormally high at this star due to magnetic reconnection events with the planet \citep{ilin2025}. We infer our flare hypothesis may be a particularly suitable explanation for this planet.

\subsection{Emerging Population-Wide Trends}
Interestingly, of the planets that lack methane, several orbit young stars or stars with evidence of flaring: HD 189733b \citep{lecavelier2012}, GJ 3470 b \citep[][]{bourrier2021}, and HIP 67522 b (Maggio et al., 2024). Direct evidence for flaring activity has not been detected at HAT-P-18, but observing data supports sunspots and magnetic activity \citep{fournier-tondreau_near-infrared_2024}. Meanwhile, the planets for which methane has been detected in abundance generally orbit relatively old stars of 2-5 Gyr: WASP-80b \citep{triaud2013,bell_methane_2023}; K2-18b \citep[][]{cloutier2019,madhusudhan_carbon-bearing_2023}; and TOI-270 d \citep[][]{gunther_super-earth_2019,holmberg_possible_2024}. Abundant methane has also been detected at LP791-18c \citep[][]{roy_diversity_2025}, which orbits a slightly younger star of 0.5 Gyr; but no evidence of star spots or magnetic activity are known to the authors.

However, not all cases of missing methane can be pointed to young or active stars. Several planets with missing methane orbit older stars (which are thought to flare less frequently than young stars; \citep[e.g.][]{feinstein2020}). These include WASP 107 b, WASP 39b, WASP 69b, and potentially HAT-P-12 b, which orbit stars of ages 3.4 Gyr \citep{mocnik2017}, 9 Gyr \citep{faedi2011}, 2 Gyr \citep[][]{anderson2014,allart_complex_2025}, and 2.5 Gyr \citep[][]{bonomo2017,ment2018}, respectively. Our flaring hypothesis may not be a suitable explanation for the missing methane for these worlds. These planets all have inflated radii compared to their masses, a problem commonly detected for close-in gas giants \citep[e.g.][]{sarkis2021,batygin2025, charbonneau_detection_2000, bodenheimer2001, batyginstevenson2010, laughlin2011, thorngren2018, guillot2002}. At WASP 107 b, it has been proposed that Ohmic heating could account for its radius inflation and heat the interior such that strong mixing from a warm methane-depleted region could result in depleted methane detections in the atmospheric spectra \citep[e.g.][]{batygin2025}. We deem it plausible that these four planets may have large $T_{int}$ from tidal or Ohmic heating. Strong vertical mixing from this warm interior would suppress methane abundances at the observing layer \citep[][]{fortney2020}. Supporting the strong mixing, a relatively large eddy diffusivity of $10^9$ $\mathrm{cm^2/s}$ was inferred at WASP 107 b by joint fitting JWST data of the NIR water feature and mid-IR silicate feature \citep[][]{huang2026}.

An exception to these categories is TOI-421 b, which orbits an older star of 9 Gyr \citep[][]{carleo2020} with no known evidence for flaring variability and is only 2.64 earth radii \citep[][]{krenn_characterisation_2024}. A low mean molecular weight with solar metallicity in a haze-free atmosphere was inferred from water features in JWST data, and tentative detections of CO and $\mathrm{SO_2}$ suggest photochemistry could influence the atmospheric composition \citep[][]{davenport_toi-421_2025}. Further modeling efforts to understand this atmosphere may be needed.

\subsection{Oxidation Prevents Haze Formation}

Methane photolysis drives the formation of soot and tholin-like hazes at Titan \citep[][]{lindal1983,lavvas2008, tomasko2008}, Pluto \citep[][]{gladstone2016,gao2017}, and warm exoplanets \citep[][]{gao_aerosol_2020, arney2017, adams2019}. This may lead one to ask: why would faster methane photolysis from flares not create thick hazes that would mute spectral features? \citet{owenmurrayclay2025} show that radiation pressure accelerates hazes at high altitudes to faster terminal velocities, prohibiting growth; this may be particularly relevant for young systems with active stars. In addition, the formation of these hazes may be highly dependent on the C:O ratio in the atmosphere. Our photochemical model shows that oxidation to CO and $\mathrm{CO_2}$ requires $\mathrm{H_2O}$ and OH, so a lower C:O ratio may favor oxidation over polymerization of hydrocarbons. Similar chemistry has been found in other modeling works \citep[][]{moses_chemical_2013,madhusudhan2012}. Furthermore, laboratory work has demonstrates that for $C:O <1$, water vapor is abundant and likely inhibits the growth of organics and aerosols \citep[][]{fleury2020}. 

\subsection{HCN}\label{sec:HCN}

We find that VULCAN overestimates the HCN abundance by orders-of-magnitude, compared to the retrieved abundance from JWST data (Section 4.3). Changes to the elemental abundance ratios help decrease HCN, but not to the retrieved amounts. In our atmosphere, the formation of HCN is fixed-nitrogen-limited, and consistently our sensitivity tests show that HCN abundance decreases linearly with decreases in N:H. Decreasing C:O to 0.1 decreased HCN by only a factor of 2 to $5\times10^{-4}$, which agrees with previous studies of the sensitivity of HCN production to C:O ratio \citep[][]{friss2026}. \citet[][]{hu_photochemistry_2021} suggests that at temperate planets $\mathrm{NH_2}$ may react with itself to form $\mathrm{N_2}$. This motivated a test where we increased N:H to determine if more nitrogen available would prefer $\mathrm{N_2}$ formation over reacting with radicals from methane photolysis. However, due to the warm temperature profile at V1298~Tau\,b compared to the worlds in \citet[][]{hu_photochemistry_2021}, we find that increasing N:H by 10 resulted in an increase in HCN mixing ratio to $4.9\times10^{-3}$. 

Predictions of large HCN abundances by photochemical models are well established in the literature for close-in planets with $\mathrm{H_2}$-rich atmospheres \citep[][]{line2011,moses_disequilibrium_2011, kawashima2018, hobbs2019, adams2022_titan}. In the solar system, HCN does not build up at Jupiter since $\mathrm{NH_3}$ and $\mathrm{CH_4}$ photolyze at different heights. At Titan, $\mathrm{N_2}$ dissociation in the upper atmosphere in the presence of $\mathrm{CH_4}$ photolysis encourages HCN formation, and HCN increases gradually with altitude to $>100s$ ppm \citep[e.g.][]{pearce2020}. Since methane and ethane are more easily photolyzed than HCN \citep[e.g.][]{ugelow2024}, HCN is thought to be relatively stable against photolysis and generally builds up in photochemical models.

Several works also report missing $NH_3$ in similar atmospheres. TOI 270 d and K2 18 b both are underabundant in $NH_3$ \citep[][]{benneke_jwst_2024, hu_water-rich_2025}, and it is plausible enhanced photolysis of $NH_3$ from stellar activity may have photochemically produced $HCN$ instead.

No detections of HCN have been made yet in an exoplanet atmosphere, though. This discrepancy between data and photochemical models has recently been noticed, and new works have begun to investigate whether nitrile reactions are missing from default chemical networks \citep[][]{veillet2024}. The key reactions involving $CH_2NH_2$ as a pathway back to $\mathrm{NH_3}$ (rather than HCN) are included in our VULCAN simulations. When updating the full nitrile work to the CHON network in \citet[][]{veillet2024}, HCN decreased by less than a factor of 2. So our nitrile chemistry is primarily up-to-date with the literature. Other works suggest sulfur chemistry may interfere with radicals from methane photolysis \citep[][]{veillet2026}, sometimes competing with and slowing HCN formation (eg, at HD 189733b), but other times providing a different route to HCN formation resulting in very similar abundances (eg, at WASP 39b). Sulfur chemistry is beyond the scope of this work, but we encourage future work to test whether including sulfur chemistry alters the HCN abundance from our results.

The primary destruction of HCN is photolysis, and due to limitations in laboratory data we ignore potential temperature dependences. This specie is difficult to work with in a laboratory setting, so photolysis cross section data is limited. The cross sections in VULCAN come from works that assumed room temperature. Ongoing research is working to measure cross sections at higher temperatures but to date none have been published \citep[][]{chubb2024} At higher temperatures, more molecules already have excited rotational or vibrational states so less photon energy is needed for electronic transitions. So, cross sections would likely increase for less energetic photons in the tail region. Since cross sections are orders of magnitude smaller here than for example at Lyman-Alpha, the temperature dependence of the total photolysis rate is likely to be small. Preliminary results shown in conference proceedings are consistent with this (Collado et al 2023, 2025).



\section{Conclusion}
We have shown that photochemistry driven by stellar flares can explain the missing methane observed at V1298~Tau\,b for nominal internal temperatures of about 100 K. This relaxes the requirement of a very high internal temperature ($\sim$ 500 K) to explain the missing methane, which is inconsistent with planet formation theories.

During repeated stellar flares, methane photolysis becomes fast enough to overcome replenishment by vertical mixing, driving the atmosphere toward a steady state methane abundance profile that is depleted with respect to that for a quiescent star.
The exact amount of depletion depends on the energy and frequency of the flares. Our model with a 15,000 K flare covering 0.3$\%$ of the stellar surface area with a $20\%$ frequency can reproduce the $\mathrm{CH_4}$, $\mathrm{CO}$, and $\mathrm{CO_2}$ JWST measurements in \citet[][]{barat_metal-poor_2025}. In addition, we also showed for comparison that equilibrium chemistry and quiescent photochemistry scenarios do not match the observations.

More broadly, our results suggest that flare-driven photochemistry may contribute to the missing methane problem across the known population of warm sub-Neptunes and gas giants, particularly those orbiting young, active stars. We discuss an emerging trend that several planets with missing methane show independent evidence of stellar flaring (e.g., HD 189733b, GJ 3470b, HIP 67522b), making them plausible candidates for this scenario. However, missing methane has also been detected at planets orbiting quieter, older stars, and they may still require alternative explanations with a high $T_{int}$. 

One discrepancy in our best-fit model is an overprediction of HCN relative to the observed spectrum, which may highlight missing reactions in existing photochemical networks. HCN build-up at other close-in planets has been hypothesized previously \citep[][]{adams2022_titan, rimmer2019} despite few HCN detections in the exoplanet literature thus far. Recent works have begun to expand the nitrile reaction networks \citep[e.g.][]{veillet2024,veillet2026}, and we encourage future modeling work to continue these investigations. Alternatively, if the presence of HCN persists in the model predictions, even when more extensive chemical networks are implemented, it may serve as an observational probe of atmospheric chemistry under flaring conditions.

\begin{acknowledgments}
We thank James Owen for insightful discussions and comments that helped improve the manuscript.
 D.A.’s research is funded by NASA through the NASA Hubble Fellowship Program Grant HST-HF2-51523.001-A awarded by the Space Telescope Science Institute, which is operated by the Association of Universities for Research in Astronomy, Inc., for NASA, under contract NAS5‐26555. H.E.S. gratefully acknowledges support from the NASA Exoplanet Research Program under grant number 80NSSC25K7143. Support for P.A.R. through programs \#5967, \#6457, \#9095, and \#12157 was provided by NASA through grants from the Space Telescope Science Institute, which is operated by the Association of Universities for Research in Astronomy, Inc., under NASA contract NAS 5-03127. This work is based in part on observations made with the NASA/ESA/CSA James Webb Space Telescope. These observations are associated with program \#2149.
\end{acknowledgments}

\appendix


\section{Toy Model Description}

The flare energy, duration, and frequency all affect the steady state mixing ratio near the observed atmospheric layer (approximately 2 mbar) in a non-linear way. To run a full photochemical model over hundreds of individual flares (as in Figure 7) involves a very long simulation run time. So to illuminate the non-liner behavior we employ a simple ``toy" model, where we approximate the methane profile as only a balance between diffusion and photolysis. Our toy model is only a good approximation in the region where photolysis and diffusion compete, which Figure 6 shows is near the layer that can be probed by observations. 

In this simple toy model, where photolysis and diffusion balance but other photochemistry is ignored, the steady state methane abundance can be written: 

\begin{equation}
    K_{zz} \frac{d^2f}{dz^2} - \frac{1}{H} \frac{df}{dz} = J f    
\end{equation}

where J is the photolysis rate coefficient ($s^{-1}$), $K_{zz}$ is the eddy diffusion coefficient ($cm^2/s$), H is the scale height, and f is the mixing ratio of methane. The solution shows that 

\begin{equation}
    f \propto (J)^{-1/4} e^{-\sqrt{J}}
\end{equation}

and in its full form is 

\begin{equation}
    f(z) = f_o \sqrt{\frac{\kappa_o}{\kappa_z}} \exp(-\int_o^z(\kappa_{z_i} - \frac{1}{2H_{z_i}})dz_i)
\end{equation}

where $f_o$ describes the mixing ratio at the bottom of the modeled atmosphere ($z_o$), $f(z)$ describes the mixing ratio of methane at altitude $z$, and 

\begin{equation}
    \kappa = \sqrt{\frac{J}{K_{zz}} + \frac{1}{4H^2}}
\end{equation}

The flare frequency, $\gamma$ (duration of a flare : total simulated duration) influences the time-averaged photolysis rate coefficient: 

\begin{equation}
J = \gamma \int{\sigma F_{flare}(\lambda) d\lambda} + (1-\gamma)  \int{\sigma F_{quiet}(\lambda) d\lambda} 
\end{equation}

where $\sigma$ describes the photolysis cross section and $F_{flare}$ and $F_{quiet}$ describe the  stellar photon flux at a wavelength $\lambda$. In this simplified model, we adopt the flare and quiescent J profiles from the steady-state Vulcan cases for flare and quiescent periods, and we adjust flare : total duration ratios by adjusting $\gamma$ and $\gamma -1$. This ignores feedback effects considered in Vulcan, where continuous flares may deplete methane and other opaque species in the upper atmosphere (e.g., $H_2O, CO_2$) that would alter $J_{flare}$ and $J_{quiet}$ over time. Figure 2 shows that changes in $J_{flare}$ and $J_{quiet}$ over time from inconsistent stellar activity primarily span the pressure levels from 100 mbar to a few mbar. This means our simplified sensitivity test over-estimates these J's early in the simulation compared to the full photochemistry in VULCAN (see Figure 2).

Ignoring subsequent chemistry after photolysis generally underestimates methane near and below the region relevant to observations, since reactions below 2 mbar can recycle the photolysis products back to CH$_4$. Our toy model also over-estimates methane in the upper atmosphere. The reason is that the simplified model ignores reactions with atomic H that otherwise would destroy methane more efficiently in this region. 

We use the toy model only to demonstrate the non-linear relationship between methane mixing ratio at 2 mbar and the flare frequency, which we visualize in Figure \ref{fig:toy}. This relationship holds as long as the flare duration is shorter than the relaxation timescale. If the flare or no-flare durations exceed the relaxation timescale, the mixing ratio will reach the steady state for that stellar spectrum and then plateau until the spectrum changes. In this case, no new steady state can be reached and the system will oscillate between the two end-member steady state cases.



\begin{figure*}
\gridline{
\fig{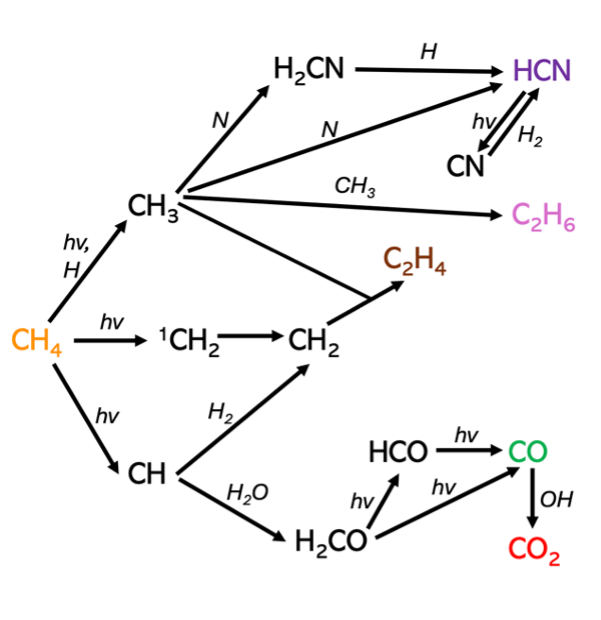}{0.4\textwidth}{}
}
\caption{Cartoon network of methane destruction and pathways to CO, CO2, HCN, and hydrocarbons.}
\label{fig:cartoon}
\end{figure*}

\begin{figure*}
\gridline{
\fig{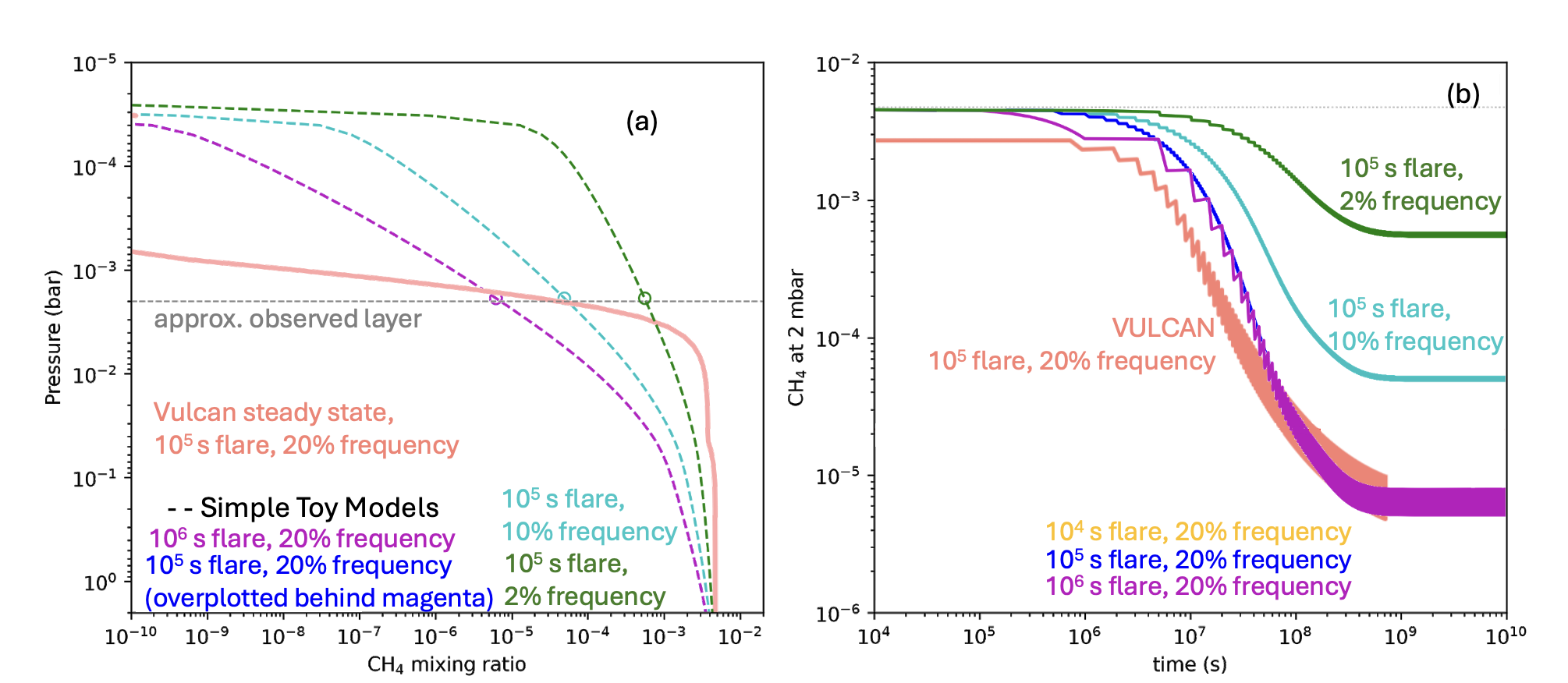}{0.9\textwidth}{}
}
\caption{Testing methane depletion for different flare frequencies and durations in the toy model. Steady state mixing ratio profiles of $\mathrm{CH_4}$ are shown from the toy model for $10^6$ s flares with 20 \% frequency (magenta), $10^5$ s flares of 20\% frequency (blue; overplotted and exactly matches the magenta profile); $10^5$ s flares with 10\% frequency (light blue); $10^5$ s flares of 2\% frequency (green). For comparison, the full photochemical result from VULCAN for $10^5$ s flares with 20\% frequency are shown in peach (this is the same flaring scenario as the toy model's blue profile). In VULCAN, in the lower atmosphere (below the observed layer), fast reactions to restore the methane under high temperatures contribute to maintaining a nearly constant methane profile. Near the observed layer, photolysis and diffusion compete, making the toy model a good assumption. Above the observed layer, methane is destroyed faster in VULCAN by reactions with atomic H (but the toy model ignores these reactions). So, the toy model results only reproduce the full photochemical model well near the observed layer where diffusion and photolysis are the main drivers of the methane sources and sinks. 
}
\label{fig:toy}
\end{figure*}




\begin{figure*}
\fig{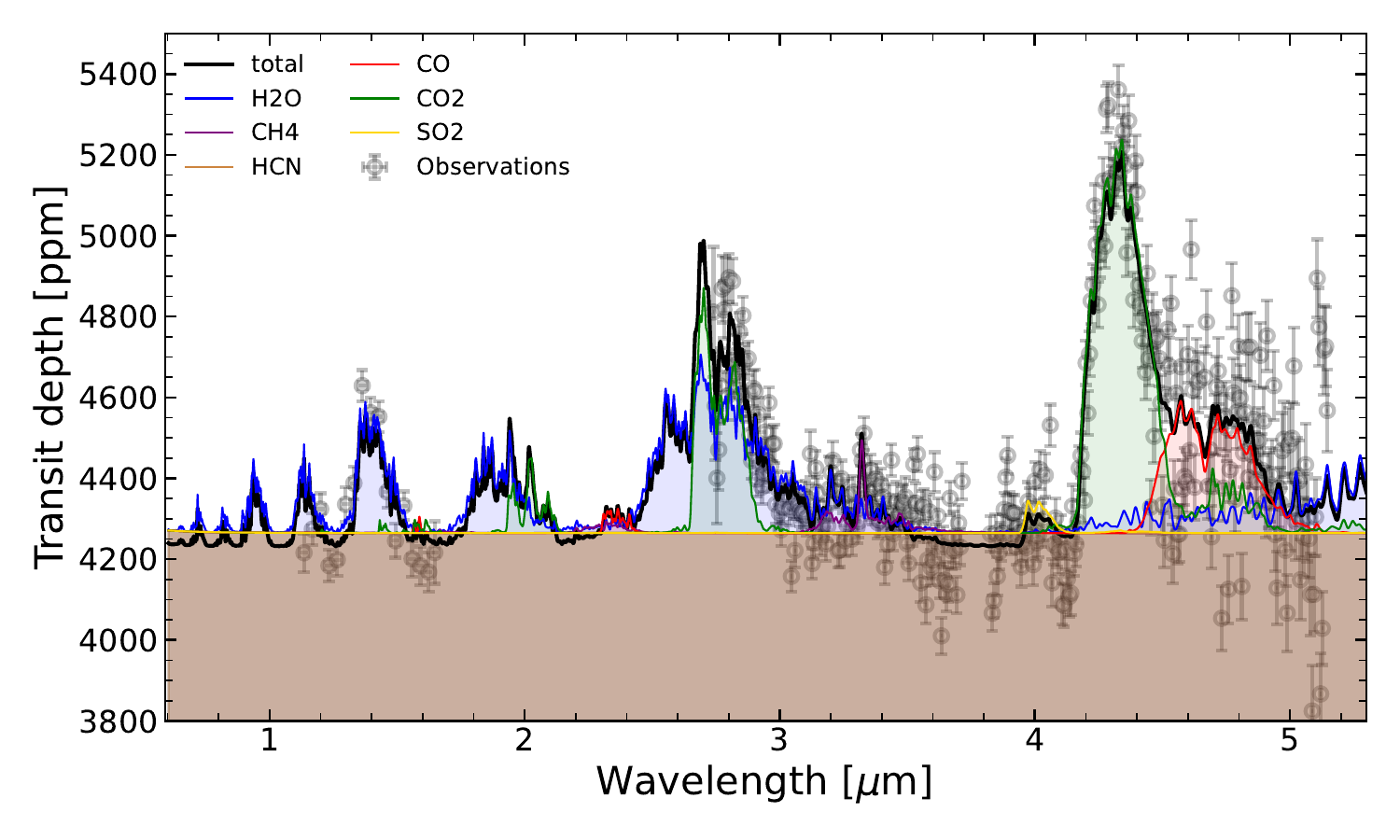}{0.85\textwidth}{}
\caption{Transmission spectrum of V1298~Tau\,b observed with HST/WFC3 and JWST/NIRSpec G395H. The transit spectrum of V1298~Tau\,b is shown as grey data points with 1$\sigma$ error bars for the HST/WFC3 transit \citep{barat_metal-poor_2024} and the JWST/NIRSpec G395H transit \citep{barat_metal-poor_2025}. The simulated transmission spectrum of the best-fit model retrieved with SCARLET is shown in black, with the molecular contributions shown as the colored regrions. The brown area represents the cloud deck opacity contribution. Water, carbon dioxyde, and carbon monoxide show clear absorption bands. The observations only show a shallow methane absorption band around the Q-branch (at 3.3\,$\mu$m). }
\label{fig:bestfitmodel}
\end{figure*}


\bibliography{refs}{}

@article{david2019v1298,
  author  = {David, Trevor J. and Petigura, Erik A. and Luger, Rodrigo and Foreman-Mackey, Daniel and Livingston, John H. and Mamajek, Eric E. and Hillenbrand, Lynne A.},
  title   = {Four Newborn Planets Transiting the Young Solar Analog V1298 Tau},
  journal = {The Astrophysical Journal Letters},
  year    = {2019},
  volume  = {885},
  pages   = {L12},
  doi     = {10.3847/2041-8213/ab4c99}
}

@article{suarezmascareno2021,
  author  = {Su{\'a}rez Mascare{\~n}o, Alejandro and Damasso, M. and Londieu, N. and Sozzetti, A. and Bejar, V. J. S. and others},
  title   = {Rapid contraction of giant planets orbiting the 20-million-year-old star V1298 Tau},
  journal = {Nature Astronomy},
  year    = {2022},
  volume  = {6},
  pages   = {232--240},
  doi     = {10.1038/s41550-021-01533-7}
}

@article{maggio2022,
  author  = {Maggio, Antonio and Locci, D. and Pillitteri, I. and Benatti, S. and Claudi, R. and others},
  title   = {New Constraints on the Future Evaporation of the Young Exoplanets in the V1298 Tau System},
  journal = {The Astrophysical Journal},
  year    = {2022},
  volume  = {925},
  pages   = {172},
  doi     = {10.3847/1538-4357/ac4040}
}

@article{finociety2023,
  author  = {Finociety, Benjamin and Donati, J. F. and Cristofari, P. I. and Moutou, C. and Cadieux, C. and others},
  title   = {Monitoring the young planet host V1298 Tau with SPIRou: planetary system and evolving large-scale magnetic field},
  journal = {Monthly Notices of the Royal Astronomical Society},
  year    = {2023},
  volume  = {526},
  pages   = {4627}
}

@article{brems2019,
  author  = {Brems, Stefan S. and K{\"u}rster, Martin and Trifonov, Trifon and Reffert, Sabine and Quirrenbach, Andreas},
  title   = {Radial-velocity jitter of stars as a function of observational timescale and stellar age},
  journal = {Astronomy \& Astrophysics},
  year    = {2019},
  volume  = {632},
  pages   = {A37},
  doi     = {10.1051/0004-6361/201935520}
}

@article{tran2024,
  author  = {Tran, Quang H. and Bowler, Brendan P. and Cochran, W. D. and Halyerson, S. and Mahadevan, S. and Ninan, Joe P. and others},
  title   = {The Epoch of Giant Planet Migration Planet Search Program. II. Documenting the Diminishing Activity of Young Suns},
  journal = {The Astronomical Journal},
  volume = {167},
  year    = {2024},
  doi     = {10.3847/1538-3881/ad2eaf},
}

@article{blunt2023,
  author  = {Blunt, Sarah and Carvalho, Adolfo and David, Trevor J. and Beichman, Charles and Zink, Jon K. and Gaidos, Eric and others},
  title   = {Overfitting Affects the Reliability of Radial Velocity Mass Estimates of the V1298 Tau Planets},
  journal = {The Astronomical Journal},
  year    = {2023},
  volume  = {166},
  pages   = {62},
  doi     = {10.3847/1538-3881/acde78}
}

@article{feinstein2021v1298,
  author  = {Feinstein, Adina D. and Montent, Benjamin T. and Johnson, Marshall C. and Bean, Jacob L. and David, Trevor D. and others},
  title   = {H-alpha and Ca II Infrared Triplet Variations During a Transit of the 23 Myr Planet V1298 Tau c},
  journal = {The Astronomical Journal},
  year    = {2021},
  volume  = {162},
  pages   = {213},
  doi     = {10.3847/1538-3881/ac1f24}
}

@article{vissapragada2021,
  author  = {Vissapragada, Shreyas and Stefansson, Gudmundur and Greklek-McKeon, M. and Oklopcic, Antonija and Knutson, Heather A. and others},
  title   = {A Search for Planetary Metastable Helium Absorption in the V1298 Tau System},
  journal = {The Astronomical Journal},
  year    = {2021},
  volume  = {162},
  pages   = {222},
  doi     = {10.3847/1538-3881/ac1bb0}
}

@article{moses2011,
  author  = {Moses, Julianne I. and Visscher, Channon and Fortney, Jonathan J. and Showman, Adam P. and Lewis, Nikole K. and Griffith, Caitlin A. and Klippenstein, Stephen J. and Shabram, Megan and Friedson, Andrew J. and Marley, Mark S. and Freedman, Richard S.},
  title   = {Disequilibrium Carbon, Oxygen, and Nitrogen Chemistry in the Atmospheres of HD 189733b and HD 209458b},
  journal = {The Astrophysical Journal},
  year    = {2011},
  volume  = {737},
  pages   = {15},
  doi     = {10.1088/0004-637X/737/1/15}
}

@article{roudier2021,
  author  = {Roudier, Gael M. and Swain, Mark R. and Gudipati, Murthy and West, Robert A. and Estrela, R. and Zellem, Robert T.},
  title   = {Disequilibrium Chemistry in Exoplanet Atmospheres Observed with the Hubble Space Telescope},
  journal = {The Astronomical Journal},
  year    = {2021},
  volume  = {162},
  pages   = {37},
  doi     = {10.3847/1538-3881/abfdad}
}

@article{baxter2021,
  author  = {Baxter, Claire and D{\'e}sert, Jean-Michel and Tsai, Shang-Min and Todorov, Kamen, O. and Bean, Jacob and Deming, Drake and Parmentier, Vivien and Fortney, Jonathan and others},
  title   = {Evidence for disequilibrium chemistry from vertical mixing in hot Jupiter atmospheres: A comprehensive survey of transiting exoplanets from 3.6 to 4.5 $\mu$m Spitzer photometry},
  journal = {Astronomy \& Astrophysics},
  year    = {2021},
  volume  = {648},
  pages   = {A127},
  doi     = {10.1051/0004-6361/202039708}
}

@article{tsai2023so2,
  author  = {Tsai, Shang-Min and Lee, Elspeth K. H. and Powell, Diana and Gao, Peter and Zhang, Xi and Moses, Julianne and Hebrard, Eric and Venot, Olivia and others},
  title   = {Photochemically produced SO$_2$ in the atmosphere of WASP-39b},
  journal = {Nature},
  year    = {2023},
  volume  = {617},
  pages   = {483--487},
  doi     = {10.1038/s41586-023-05902-2}
}

@article{wakeford2018,
  author  = {Wakeford, Hannah R. and Sing, D. and Deming, D. and Lewis, N. and Goyal, J. and Wilson, T. and others},
  title   = {The Complete Transmission Spectrum of WASP-39b with a Precise Water Constraint},
  journal = {The Astronomical Journal},
  year    = {2018},
  volume  = {155},
  pages   = {29},
  doi     = {10.3847/1538-3881/aa9e4e}
}

@article{chachan2019,
  author  = {Chachan, Yayaati and Knutson, Heather A. and Gao, P. and Kataria, T. and Wong, I. and others},
  title   = {A Hubble PanCET Study of HAT-P-11b: A Cloudy Neptune with a Low Atmospheric Metallicity},
  journal = {The Astronomical Journal},
  year    = {2019},
  volume  = {158},
  pages   = {244},
  doi     = {10.3847/1538-3881/ab4e9a}
}

@article{benneke2019gj3470,
  author  = {Benneke, Bj{\"o}rn and Knutson, Heather A. and Lothringer, Joshua and Crossfield, Ian J. M. and Moses, Julianne I. and Morley, Caroline and others},
  title   = {A sub-Neptune exoplanet with a low-metallicity methane-depleted atmosphere and Mie-scattering clouds},
  journal = {Nature Astronomy},
  year    = {2019},
  volume  = {3},
  pages   = {813--821},
  doi     = {10.1038/s41550-019-0800-5}
}

@article{spake2018,
  author  = {Spake, Jessica J. and Sing, David K. and Evans, Thomas M. and Oklopcic, A. and Bourrier, V. and Kreidberg, L. and others},
  title   = {Helium in the eroding atmosphere of an exoplanet},
  journal = {Nature},
  year    = {2018},
  volume  = {557},
  pages   = {68--70},
  doi     = {10.1038/s41586-018-0067-5}
}

@article{carone2021,
  author  = {Carone, Ludmila and Molliere, Paul and Zhou, Yifan and Bouwman, Jeroen and Yan, Fei and others},
  title   = {Indications for very high metallicity and absence of methane in the eccentric exo-Saturn WASP-117b},
  journal = {Astronomy \& Astrophysics},
  year    = {2021},
  volume  = {646},
  pages   = {A168},
  doi     = {10.1051/0004-6361/202038620}
}

@article{beatty2024,
  author  = {Beatty, Thomas G. and Welbanks, Luis and Schlawin, Everett and Bell, Taylor and Line, Michael and Murphy, Matthew and others},
  title   = {Sulfur Dioxide and Other Molecular Species in the Atmosphere of the Sub-Neptune GJ 3470 b},
  journal = {The Astrophysical Journal Letters},
  year    = {2024},
  volume  = {970},
  pages   = {L10},
  doi     = {10.3847/2041-8213/ad55e9}
}

@article{knutson2011,
  author  = {Knutson, Heather A. and Madhusudhan, Nikku and Cowan, Nicolas B. and Christiansen, J. and Agol, E. and others},
  title   = {A Spitzer Transmission Spectrum for the Exoplanet GJ 436b, Evidence for Stellar Variability, and Constraints on Dayside Flux Variations},
  journal = {The Astrophysical Journal},
  year    = {2011},
  volume  = {735},
  pages   = {27},
  doi     = {10.1088/0004-637X/735/1/27}
}

@article{lanotte2014,
  author  = {Lanotte, Audrey A. and Gillon, Micha{\"e}l and Demory, Brice-Olivier and others},
  title   = {A global analysis of Spitzer and new HARPS data confirms the loneliness and metal-richness of GJ 436 b},
  journal = {Astronomy \& Astrophysics},
  year    = {2014},
  volume  = {572},
  pages   = {A73},
  doi     = {10.1051/0004-6361/201424373}
}

@article{parmentier2013,
  author  = {Parmentier, Vivien and Showman, Adam P. and Lian, Yuan},
  title   = {3D mixing in hot Jupiters atmospheres. I. Application to the day/night cold trap in HD 209458b},
  journal = {Astronomy \& Astrophysics},
  year    = {2013},
  volume  = {558},
  pages   = {A91},
  doi     = {10.1051/0004-6361/201321132}
}

@article{ackerman2001,
  author  = {Ackerman, Andrew S. and Marley, Mark S.},
  title   = {Precipitating Condensation Clouds in Substellar Atmospheres},
  journal = {The Astrophysical Journal},
  year    = {2001},
  volume  = {556},
  pages   = {872--884},
  doi     = {10.1086/321540}
}

@article{liu2026kzz,
  author  = {Liu, Jingyu and Christie, Duncan and Yang, Jun},
  title   = {Three-dimensional transport-induced chemistry on temperate sub-Neptune K2-18b, Part II: the combined effects of atmospheric dynamics and chemical reactions},
  journal = {Monthly Notices of the Royal Astronomical Society},
  year    = {2026},
  volume  = {549},
  doi     = {10.1093/mnras/stag780},
  eprint  = {2604.07987},
  archivePrefix = {arXiv}
}

@article{fortney2020,
  author  = {Fortney, Jonathan J. and Visscher, Channon and Marley, Mark S. and Hood, Callie E. and Line, Michael R. and Thorngren, Daniel P. and Freedman, Richard S. and Lupu, Roxana},
  title   = {Beyond Equilibrium Temperature: How the Atmosphere/Interior Connection Affects the Onset of Methane, Ammonia, and Clouds in Warm Transiting Giant Planets},
  journal = {The Astronomical Journal},
  year    = {2020},
  volume  = {160},
  pages   = {288},
  doi     = {10.3847/1538-3881/abc5bd}
}

@article{guillot2002,
  author  = {Guillot, Tristan and Showman, Adam P.},
  title   = {Evolution of ``51 Pegasus b-like'' planets},
  journal = {Astronomy \& Astrophysics},
  year    = {2002},
  volume  = {385},
  pages   = {156--165},
  doi     = {10.1051/0004-6361:20011624}
}

@article{yu2026,
  author  = {Yu, Xinting and Glein, Christopher R. and Thorngren, Daniel P. and Murray, David F.},
  title   = {Unusually Hot Interiors Could Reconcile the Missing-methane Problem for Warm-to-hot Exoplanets with Hydrogen Atmospheres},
  journal = {The Astrophysical Journal},
  year    = {2026},
  doi     = {10.3847/1538-4357/ae5b6b},
  eprint  = {2604.01672},
  archivePrefix = {arXiv}
}

@article{millholland2020,
  author  = {Millholland, Sarah and Petigura, Erik and Batygin, Konstantin},
  title   = {Tidal Inflation Reconsidered: Improved Estimates of WASP-107b's Interior Structure via Obliquity Tides},
  journal = {The Astrophysical Journal},
  year    = {2020},
  volume  = {897},
  pages   = {7},
  doi     = {10.3847/1538-4357/ab959c}
}

@article{agundez2014,
  author  = {Ag{\'u}ndez, Marcelino and Parmentier, Vivien and Venot, Olivia and Hersant, Franck and Selsis, Franck},
  title   = {Pseudo 2D chemical model of hot-Jupiter atmospheres: application to HD 209458b and HD 189733b},
  journal = {Astronomy \& Astrophysics},
  year    = {2014},
  volume  = {564},
  pages   = {A73},
  doi     = {10.1051/0004-6361/201322895}
}

@article{morley2017,
  author  = {Morley, Caroline V. and Knutson, Heather and Line, Michael and Fortney, Jonathan J. and Thorngren, Daniel and Marley, Mark S. and Teal, Dillon and Lupu, Roxana},
  title   = {Forward and Inverse Modeling of the Emission and Transmission Spectrum of GJ 436b: Investigating Metal Enrichment, Tidal Heating, and Clouds},
  journal = {The Astronomical Journal},
  year    = {2017},
  volume  = {153},
  pages   = {86},
  doi     = {10.3847/1538-3881/153/2/86}
}

@article{batygin2025,
  author  = {Batygin, Konstantin},
  title   = {From Tides to Currents: Unraveling the Mechanism that Powers WASP-107b's Internal Heat Flux},
  journal = {The Astrophysical Journal},
  year    = {2025},
  doi     = {10.3847/1538-4357/adccc4},
  eprint  = {2505.01581},
  archivePrefix = {arXiv}
}

@article{zahnle2014,
  author  = {Zahnle, Kevin J. and Marley, Mark S.},
  title   = {Methane, Carbon Monoxide, and Ammonia in Brown Dwarfs and Self-Luminous Giant Planets},
  journal = {The Astrophysical Journal},
  year    = {2014},
  volume  = {797},
  pages   = {41},
  doi     = {10.1088/0004-637X/797/1/41}
}

@article{line2010,
  author  = {Line, Michael R. and Liang, Mao-Chang and Yung, Yuk L.},
  title   = {High-temperature Photochemistry in the Atmosphere of HD 189733b},
  journal = {The Astrophysical Journal},
  year    = {2010},
  volume  = {717},
  pages   = {496--502},
  doi     = {10.1088/0004-637X/717/1/496}
}

@article{jaziri2025,
  author  = {Jaziri, A. Y. and Sohier, O. and Venot, O. and Carrasco, N.},
  title   = {Unraveling the non-equilibrium chemistry of the temperate sub-Neptune K2-18 b},
  journal = {Astronomy \& Astrophysics},
  year    = {2025},
  volume  = {701},
  pages   = {A33},
  doi = {10.1051/0004-6361/202555496}
}

@article{kempton2012,
  author  = {Miller-Ricci Kempton, Eliza and Zahnle, Kevin and Fortney, Jonathan J.},
  title   = {The Atmospheric Chemistry of GJ 1214b: Photochemistry and Clouds},
  journal = {The Astrophysical Journal},
  year    = {2012},
  volume  = {745},
  pages   = {3},
  doi     = {10.1088/0004-637X/745/1/3}
}

@article{feinstein2020,
  author  = {Feinstein, Adina D. and Montet, Benjamin T. and Ansdell, Megan and Nord, Brian and Bean, Jacob L. and others},
  title   = {Flare Statistics for Young Stars from a Convolutional Neural Network Analysis of TESS Data},
  journal = {The Astronomical Journal},
  year    = {2020},
  volume  = {160},
  pages   = {219},
  doi     = {10.3847/1538-3881/abac0a}
}

@article{feinstein2024,
  author  = {Feinstein, Adina D. and Seligman, Darryl Z. and France, Kevin and Gagn{\'e}, Jonathan and Kowalski, Adam},
  title   = {Evolution of Flare Activity in GKM Stars Younger Than 300 Myr over Five Years of TESS Observations},
  journal = {The Astronomical Journal},
  year    = {2024},
  volume  = {168},
  pages   = {60},
  doi     = {10.3847/1538-3881/ad4edf}
}

@article{howard2020,
  author  = {Howard, Ward S. and Corbett, Hank and Law, Nicholas M. and others},
  title   = {EvryFlare. III. Temperature Evolution and Habitability Impacts of Dozens of Superflares Observed Simultaneously by Evryscope and TESS},
  journal = {The Astrophysical Journal},
  year    = {2020},
  volume  = {902},
  pages   = {115},
  doi     = {10.3847/1538-4357/abb5b4}
}

@article{paudel2024,
doi = {10.3847/1538-4357/ad487d},
url = {https://doi.org/10.3847/1538-4357/ad487d},
year = {2024},
month = {aug},
publisher = {The American Astronomical Society},
volume = {971},
number = {1},
pages = {24},
author = {Paudel, Rishi R. and Barclay, Thomas and Youngblood, Allison and Quintana, Elisa V. and Schlieder, Joshua E. and Vega, Laura D. and Gilbert, Emily A. and Osten, Rachel A. and Peacock, Sarah and Tristan, Isaiah I. and Feliz, Dax L. and Boyd, Patricia T. and Davenport, James R. A. and Huber, Daniel and Kowalski, Adam F. and Monsue, Teresa and Silverstein, Michele L.},
title = {A Multiwavelength Survey of Nearby M Dwarfs: Optical and Near-ultraviolet Flares and Activity with Contemporaneous TESS, Kepler/K2, Swift, and HST Observations},
journal = {The Astrophysical Journal},
}

@article{kowalski2024review,
  author  = {Kowalski, Adam F.},
  title   = {Stellar flares},
  journal = {Living Reviews in Solar Physics},
  year    = {2024},
  volume  = {21},
  pages   = {1},
  doi     = {10.1007/s41116-024-00039-4}
}

@article{kowalski2024megaflares,
  author  = {Kowalski, Adam F. and Osten, Rachel A. and Notsu, Yuta and Tristan, Isaiah I. and Segura, Antigona and Maehara, Hiroyuki and Namekata, Kosuke and Inoue, Shun},
  title   = {Rising Near-ultraviolet Spectra in Stellar Megaflares},
  journal = {The Astrophysical Journal},
  year    = {2024},
  volume = {978},
  doi     = {10.3847/1538-4357/ad9395},
}

@article{berger2024,
  author  = {Berger, Vera L. and Hinkle, Jason and Tucker, Michael and Shappee, Benjamin and van Saders, Jennifer and others},
  title   = {Stellar flares are far-ultraviolet luminous},
  journal = {Monthly Notices of the Royal Astronomical Society},
  year    = {2024},
  volume  = {532},
  pages   = {4436},
  doi = {10.1093/mnras/stae1648},
}

@article{ilin2025,
  author  = {Ilin, Ekaterina and Vedantham, Harish and Poppenhager, Katia and Bloot, Sanne and Callingham, Joseph and others},
  title   = {Close-in planet induces flares on its host star},
  journal = {Nature},
  year    = {2025},
  doi     = {10.1038/s41586-025-09236-z},
  volume = {643}
}

@article{zhang2025,
  author  = {Zhang, Michael and Paragas, Kimberly and Bean, Jacob L. and Yeung, Joseph and Chachan, Yayaati and Greene, Thomas P. and Lunine, Jonathan and Deming, Drake},
  title   = {Retrievals on NIRCam Transmission and Emission Spectra of HD 189733b with PLATON 6, a GPU Code for the JWST Era},
  journal = {The Astronomical Journal},
  year    = {2025},
  volume = {169},
  doi     = {10.3847/1538-3881/ad8cd2},
}

@article{thao2024,
  author  = {Thao, Pa Chia and Mann, Andrew W. and Feinstein, Adina and Gao, Peter and Thorngren, Daniel and Rotman, Yoav and others},
  title   = {The Featherweight Giant: Unraveling the Atmosphere of a 17 Myr Planet with JWST},
  journal = {The Astronomical Journal},
  year    = {2024},
  volume  = {168},
  pages   = {297},
  doi     = {10.3847/1538-3881/ad81d7}
}

@article{crouzet2025,
  author  = {Crouzet, Nicolas and Edwards, Billy and Konings, Thomas and Bouwman, J. and Min M. and Lagage, P.-O. and others},
  title   = {Detection of CO$_2$, CO, and H$_2$O in the atmosphere of the warm sub-Saturn HAT-P-12 b},
  journal = {Astronomy \& Astrophysics},
  year    = {2025},
  volume = {703},
  doi     = {10.1051/0004-6361/202450690}
}

@article{fu2022,
  author  = {Fu, Guangwei and Espinoza, Nestor and Sing, David and Lothringer, Joshua and Dos Santos, Leonardo A. and others},
  title   = {Water and an Escaping Helium Tail Detected in the Hazy and Methane-depleted Atmosphere of HAT-P-18b from JWST NIRISS/SOSS},
  journal = {The Astrophysical Journal Letters},
  year    = {2022},
  volume  = {940},
  pages   = {L35},
  doi     = {10.3847/2041-8213/ac9977}
}

@article{schlawin2024,
  author  = {Schlawin, Everett and Ohno, Kazumasa and Bell, Taylor and Murphy, Matthew and Welbanks, Luis and others},
  title   = {Possible Carbon Dioxide above the Thick Aerosols of GJ 1214 b},
  journal = {The Astrophysical Journal Letters},
  year    = {2024},
  volume={974},
  doi={10.3847/2041-8213/ad7fef}
}

@article{kreidberg2018,
  author  = {Kreidberg, Laura and Line, Michael R. and Thorngren, Daniel and Morley, Caroline V. and Stevenson, Kevin B.},
  title   = {Water, High-altitude Condensates, and Possible Methane Depletion in the Atmosphere of the Warm Super-Neptune WASP-107b},
  journal = {The Astrophysical Journal Letters},
  year    = {2018},
  volume  = {858},
  pages   = {L6},
  doi     = {10.3847/2041-8213/aabfce}
}

@article{dyrek2024,
  author  = {Dyrek, Achr{\`e}ne and Min, Michiel and Decin, Leen and Bouwman, Jeroen and Crouzet, Nocolas and Molliere, Paul and others},
  title   = {SO$_2$, silicate clouds, but no CH$_4$ detected in a warm Neptune with JWST MIRI},
  journal = {Nature},
  year    = {2024},
  volume  = {625},
  pages   = {51--54},
  doi     = {10.1038/s41586-023-06849-0}
}

@article{davenport2025,
  author  = {Davenport, Brian and Kempton, Eliza M.-R. and Nixon, Matthew C. and Bean, Jacob L. and others},
  title   = {TOI-421 b: A Hot Sub-Neptune with a Haze-free, Low Mean Molecular Weight Atmosphere},
  journal = {The Astrophysical Journal Letters},
  year    = {2025},
  volume  = {984},
  pages   = {L44},
  doi     = {10.3847/2041-8213/adcd76}
}

@article{crossfield2013,
  author  = {Crossfield, Ian J. M. and Barman, Travis and Hansen, Brad and Howard, Andrew},
  title   = {Warm ice giant GJ 3470b. I. A flat transmission spectrum indicates a hazy, low-methane, and/or metal-rich atmosphere},
  journal = {Astronomy \& Astrophysics},
  year    = {2013},
  volume  = {559},
  pages   = {A33},
  doi     = {10.1051/0004-6361/201322278}
}

@article{biddle2014,
  author  = {Biddle, Lauren I. and Pearson, Kyle A. and Crossfield, Ian and Fulton, Benjamin and Ciceri, Simona and others},
  title   = {Warm ice giant GJ 3470b -- II. Revised planetary and stellar parameters from optical to near-infrared transit photometry},
  journal = {Monthly Notices of the Royal Astronomical Society},
  year    = {2014},
  volume  = {443},
  pages   = {1810--1820},
  doi     = {10.1093/mnras/stu1199}
}

@article{bourrier2021,
  author  = {Bourrier, Vincent and dos Santos, L. A. and Sanz-Forcada, J. and Garcia Munoz, A. and Henry, G. W. and others},
  title   = {The Hubble PanCET program: long-term chromospheric evolution and flaring activity of the M dwarf host GJ 3470},
  journal = {Astronomy \& Astrophysics},
  year    = {2021},
  volume={650},
  doi={10.1051/0004-6361/202140487}
}

@article{swain2009,
  author  = {Swain, Mark R. and Vasisht, Gautam and Tinetti, Giovanna and Bouwman, Jeroen and Chen, Pin and Yung, Yuk and Deming, Drake and Deroo, Pieter},
  title   = {Molecular Signatures in the Near-infrared Dayside Spectrum of HD 189733b},
  journal = {The Astrophysical Journal Letters},
  year    = {2009},
  volume  = {690},
  pages   = {L114--L117},
  doi     = {10.1088/0004-637X/690/2/L114}
}

@ARTICLE{adams2022_titan,
       author = {{Adams}, Danica and {Luo}, Yangcheng and {Yung}, Yuk L.},
        title = "{Hydrocarbon chemistry in the atmosphere of a Warmer Exo-Titan}",
      journal = {Frontiers in Astronomy and Space Sciences},
         year = 2022,
        month = sep,
       volume = {9},
          eid = {823227},
        pages = {823227},
          doi = {10.3389/fspas.2022.823227},
       adsurl = {https://ui.adsabs.harvard.edu/abs/2022FrASS...9.3227A}
}

@ARTICLE{line2011,
       author = {{Line}, Michael R. and {Vasisht}, Gautam and {Chen}, Pin and {Angerhausen}, D. and {Yung}, Yuk L.},
        title = "{Thermochemical and Photochemical Kinetics in Cooler Hydrogen-dominated Extrasolar Planets: A Methane-poor GJ436b?}",
      journal = {\apj},
         year = 2011,
        month = sep,
       volume = {738},
       number = {1},
          eid = {32},
        pages = {32},
          doi = {10.1088/0004-637X/738/1/32},
archivePrefix = {arXiv},
       eprint = {1104.3183},
 primaryClass = {astro-ph.EP},
       adsurl = {https://ui.adsabs.harvard.edu/abs/2011ApJ...738...32L}
}

@article{gibson2011,
  author  = {Gibson, Neale P. and Pont, Fr{\'e}d{\'e}ric and Aigrain, Suzanne},
  title   = {A new look at NICMOS transmission spectroscopy of HD 189733, GJ-436 and XO-1: no conclusive evidence for molecular features},
  journal = {Monthly Notices of the Royal Astronomical Society},
  year    = {2011},
  volume  = {411},
  pages   = {2199--2213},
  doi     = {10.1111/j.1365-2966.2010.17837.x}
}

@article{lecavelier2012,
  author  = {Lecavelier des Etangs, Alain and Bourrier, Vincent and Wheatley, Peter J. and others},
  title   = {Temporal variations in the evaporating atmosphere of the exoplanet HD 189733b},
  journal = {Astronomy \& Astrophysics},
  year    = {2012},
  volume  = {543},
  pages   = {L4},
  doi     = {10.1051/0004-6361/201219363}
}

@article{rizzuto2020,
  author  = {Rizzuto, Aaron C. and Newton, Elisabeth R. and Mann, Andrew W. and others},
  title   = {TESS Hunt for Young and Maturing Exoplanets (THYME). II. A 17 Myr Old Transiting Hot Jupiter in the Sco-Cen Association},
  journal = {The Astronomical Journal},
  year    = {2020},
  volume  = {160},
  pages   = {33},
  doi     = {10.3847/1538-3881/ab94b7}
}

@article{maggio2024,
  author  = {Maggio, Antonio and Pillitteri, I. and Argiroffi, C. and Locci, D. and Benatti, S. and Micela, G.},
  title   = {XUV irradiation of young planetary atmospheres. Results from a
joint XMM-Newton and HST observation of HIP67522},
  journal = {Astronomy \& Astrophysics},
  year    = {2024},
  volume={690},
  doi={10.1051/0004-6361/202451582}
}

@ARTICLE{piaulet2023,
       author = {{Piaulet}, Caroline and {Benneke}, Bj{\"o}rn and {Almenara}, Jose M. and {Dragomir}, Diana and {Knutson}, Heather A. and {Thorngren}, Daniel and {Peterson}, Merrin S. and {Crossfield}, Ian J.~M. and {Kempton}, Eliza M.-R. and {Kubyshkina}, Daria and {Howard}, Andrew W. and {Angus}, Ruth and {Isaacson}, Howard and {Weiss}, Lauren M. and {Beichman}, Charles A. and {Fortney}, Jonathan J. and {Fossati}, Luca and {Lammer}, Helmut and {McCullough}, P.~R. and {Morley}, Caroline V. and {Wong}, Ian},
        title = "{Evidence for the volatile-rich composition of a 1.5-Earth-radius planet}",
      journal = {Nature Astronomy},
         year = 2023,
        month = feb,
       volume = {7},
        pages = {206-222},
          doi = {10.1038/s41550-022-01835-4},
archivePrefix = {arXiv},
       eprint = {2212.08477},
 primaryClass = {astro-ph.EP},
       adsurl = {https://ui.adsabs.harvard.edu/abs/2023NatAs...7..206P}
}

@ARTICLE{livingston2026,
       author = {{Livingston}, John H. and {Petigura}, Erik A. and {David}, Trevor J. and {Masuda}, Kento and {Owen}, James and {Nesvorn{\'y}}, David and {Batygin}, Konstantin and {de Leon}, Jerome and {Mori}, Mayuko and {Ikuta}, Kai and {Fukui}, Akihiko and {Watanabe}, Noriharu and {Orell Miquel}, Jaume and {Murgas}, Felipe and {Parviainen}, Hannu and {Korth}, Judith and {Libotte}, Florence and {Abreu Garc{\'\i}a}, N{\'e}stor and {Gallardo}, Pedro Pablo Meni and {Narita}, Norio and {Pall{\'e}}, Enric and {Tamura}, Motohide and {Yonehara}, Atsunori and {Ridden-Harper}, Andrew and {Bieryla}, Allyson and {Trani}, Alessandro A. and {Mamajek}, Eric E. and {Ciardi}, David R. and {Gorjian}, Varoujan and {Hillenbrand}, Lynne A. and {Rebull}, Luisa M. and {Newton}, Elisabeth R. and {Mann}, Andrew W. and {Vanderburg}, Andrew and {Stef{\'a}nsson}, Gu{\dh}mundur and {Mahadevan}, Suvrath and {Ca{\~n}as}, Caleb and {Ninan}, Joe and {Higuera}, Jesus and {Todorov}, Kamen and {D{\'e}sert}, Jean-Michel and {Pino}, Lorenzo},
        title = "{A young progenitor for the most common planetary systems in the Galaxy}",
      journal = {\nat},
         year = 2026,
        month = jan,
       volume = {649},
       number = {8096},
        pages = {310-314},
          doi = {10.1038/s41586-025-09840-z},
archivePrefix = {arXiv},
       eprint = {2601.10598},
 primaryClass = {astro-ph.EP},
       adsurl = {https://ui.adsabs.harvard.edu/abs/2026Natur.649..310L}
}

@ARTICLE{owenmurrayclay2025,
       author = {{Owen}, James E. and {Murray-Clay}, Ruth A.},
        title = "{Aerosol dynamics on hot exoplanets: the role of radiation pressure}",
      journal = {\mnras},
         year = 2025,
        month = oct,
       volume = {543},
       number = {1},
        pages = {587-607},
          doi = {10.1093/mnras/staf1403},
archivePrefix = {arXiv},
       eprint = {2508.20175},
 primaryClass = {astro-ph.EP},
       adsurl = {https://ui.adsabs.harvard.edu/abs/2025MNRAS.543..587O}
}

@ARTICLE{chubb2024,
       author = {{Chubb}, Katy L. and {Robert}, S{\'e}verine and {Sousa-Silva}, Clara and {Yurchenko}, Sergei N. and {Allard}, Nicole F. and {Boudon}, Vincent and {Buldyreva}, Jeanna and {Bultel}, Benjamin and {Coustenis}, Athena and {Foltynowicz}, Aleksandra and {Gordon}, Iouli E. and {Hargreaves}, Robert J. and {Helling}, Christiane and {Hill}, Christian and {Hrodmarsson}, Helgi Rafn and {Karman}, Tijs and {Lecoq-Molinos}, Helena and {Migliorini}, Alessandra and {Rey}, Micha{\"e}l and {Richard}, Cyril and {Sadiek}, Ibrahim and {Schmidt}, Fr{\'e}d{\'e}ric and {Sokolov}, Andrei and {Stefani}, Stefania and {Tennyson}, Jonathan and {Venot}, Olivia and {Wright}, Sam O.~M. and {Arenales-Lope}, Rosa and {Barstow}, Joanna K. and {Bocchieri}, Andrea and {Carrasco}, Nathalie and {Dubey}, Dwaipayan and {Egorov}, Oleg and {Mu{\~n}oz}, Antonio Garc{\'\i}a and {Gharib-Nezhad}, Ehsan (Sam) and {Gkouvelis}, Leonardos and {Gr{\"u}bel}, Fabian and {Irwin}, Patrick Gerard Joseph and {Kn{\'\i}{\v{z}}ek}, Anton{\'\i}n and {Lewis}, David A. and {Lodge}, Matt G. and {Ma}, Sushuang and {Martins}, Zita and {Molaverdikhani}, Karan and {Morello}, Giuseppe and {Nikitin}, Andrei and {Panek}, Emilie and {Rengel}, Miriam and {Rinaldi}, Giovanna and {Skinner}, Jack W. and {Tinetti}, Giovanna and {van Kempen}, Tim A. and {Yang}, Jingxuan and {Zingales}, Tiziano},
        title = "{Data availability and requirements relevant for the Ariel space mission and other exoplanet atmosphere applications}",
      journal = {RAS Techniques and Instruments},
         year = 2024,
        month = jan,
       volume = {3},
       number = {1},
        pages = {636-690},
          doi = {10.1093/rasti/rzae039},
archivePrefix = {arXiv},
       eprint = {2404.02188},
 primaryClass = {astro-ph.IM},
       adsurl = {https://ui.adsabs.harvard.edu/abs/2024RASTI...3..636C}
}

@ARTICLE{maehara2015,
       author = {{Maehara}, Hiroyuki and {Shibayama}, Takuya and {Notsu}, Yuta and {Notsu}, Shota and {Honda}, Satoshi and {Nogami}, Daisaku and {Shibata}, Kazunari},
        title = "{Statistical properties of superflares on solar-type stars based on 1-min cadence data}",
      journal = {Earth, Planets and Space},
         year = 2015,
        month = dec,
       volume = {67},
          eid = {59},
        pages = {59},
          doi = {10.1186/s40623-015-0217-z},
archivePrefix = {arXiv},
       eprint = {1504.00074},
 primaryClass = {astro-ph.SR},
       adsurl = {https://ui.adsabs.harvard.edu/abs/2015EP&S...67...59M}
}

@article{doyle2020,
    author = {Doyle, L and Ramsay, G and Doyle, J G},
    title = {Superflares and variability in solar-type stars with TESS in the Southern hemisphere},
    journal = {Monthly Notices of the Royal Astronomical Society},
    volume = {494},
    number = {3},
    pages = {3596-3610},
    year = {2020},
    month = {05},
    issn = {0035-8711},
    doi = {10.1093/mnras/staa923},
    url = {https://doi.org/10.1093/mnras/staa923},
    eprint = {https://academic.oup.com/mnras/article-pdf/494/3/3596/33145067/staa923.pdf},
}

@article{liu1984,
  author  = {Liu, Shaw C. and McAfee, James R. and Cicerone, Ralph J.},
  title   = {Radon 222 and Tropospheric Vertical Transport},
  journal = {Journal of Geophysical Research: Atmospheres},
  year    = {1984},
  volume  = {89},
  number  = {D5},
  pages   = {7291--7297},
  doi     = {10.1029/JD089iD05p07291}
}

@ARTICLE{woo1981,
       author = {{Woo}, R. and {Ishimaru}, A.},
        title = "{Eddy diffusion coefficient for the atmosphere of Venus from radio scintillation measurements}",
      journal = {\nat},
         year = 1981,
        month = jan,
       volume = {289},
        pages = {383},
          doi = {10.1038/289383a0},
       adsurl = {https://ui.adsabs.harvard.edu/abs/1981Natur.289..383W}
}

@ARTICLE{thao2023,
       author = {{Thao}, Pa Chia and {Mann}, Andrew W. and {Gao}, Peter and {Owens}, Dylan A. and {Vanderburg}, Andrew and {Newton}, Elisabeth R. and {Tang}, Yao and {Fields}, Matthew J. and {David}, Trevor J. and {Irwin}, Jonathan M. and {Husser}, Tim-Oliver and {Charbonneau}, David and {Ballard}, Sarah},
        title = "{Hazy with a Chance of Star Spots: Constraining the Atmosphere of Young Planet K2-33b}",
      journal = {\aj},
         year = 2023,
        month = jan,
       volume = {165},
       number = {1},
          eid = {23},
        pages = {23},
          doi = {10.3847/1538-3881/aca07a},
archivePrefix = {arXiv},
       eprint = {2211.07728},
 primaryClass = {astro-ph.EP},
       adsurl = {https://ui.adsabs.harvard.edu/abs/2023AJ....165...23T}
}

@ARTICLE{swain2014,
       author = {{Swain}, Mark R. and {Line}, Michael R. and {Deroo}, Pieter},
        title = "{On the Detection of Molecules in the Atmosphere of HD 189733b Using HST NICMOS Transmission Spectroscopy}",
      journal = {\apj},
         year = 2014,
        month = apr,
       volume = {784},
       number = {2},
          eid = {133},
        pages = {133},
          doi = {10.1088/0004-637X/784/2/133},
archivePrefix = {arXiv},
       eprint = {1401.7601},
 primaryClass = {astro-ph.EP},
       adsurl = {https://ui.adsabs.harvard.edu/abs/2014ApJ...784..133S}
}

@article{triaud2013,
  author  = {Triaud, Amaury H. M. J. and Anderson, David R. and Collier Cameron, Andrew and Doyle, A. P. and Fumel, A. and others},
  title   = {WASP-80b: a gas giant transiting a cool dwarf},
  journal = {Astronomy \& Astrophysics},
  year    = {2013},
  volume  = {551},
  pages   = {A80},
  doi     = {10.1051/0004-6361/201220900}
}

@article{cloutier2019,
  author  = {Cloutier, Ryan and Astudillo-Defru, N. and Doyon, R. and Bonfils, X. and Almenara, J.-M. and Bouchy, F. and others},
  title   = {Confirmation of the radial velocity super-Earth K2-18 c with HARPS and CARMENES},
  journal = {Astronomy \& Astrophysics},
  year    = {2019},
  volume  = {621},
  pages   = {A49},
  doi     = {10.1051/0004-6361/201833995}
}

@article{mocnik2017,
  author  = {Mo{\v c}nik, Teo and Hellier, C. and Anderson, D. R. and Clark, B. J. M. and Southworth, J.},
  title   = {Starspots on WASP-107 and pulsations of WASP-118},
  journal = {Monthly Notices of the Royal Astronomical Society},
  year    = {2017},
  volume  = {469},
  pages   = {1622--1629},
  doi     = {10.1093/mnras/stx972}
}

@article{faedi2011,
  author  = {Faedi, Francesca and Barros, Susana C. C. and Anderson, David R. and Brown, D. J. A. and Cameron, A. Collier and Pollacco, D. and others},
  title   = {WASP-39b: a highly inflated Saturn-mass planet orbiting a late G-type star},
  journal = {Astronomy \& Astrophysics},
  year    = {2011},
  volume  = {531},
  pages   = {A40},
  doi     = {10.1051/0004-6361/201116671}
}

@article{anderson2014,
  author  = {Anderson, David R. and Cameron, A. Collier and Delrez, L. and Doyle, A. P. and Faedi, F. and Fumel, A. and others},
  title   = {Three newly discovered sub-Jupiter-mass planets: WASP-69b and WASP-84b transit active K dwarfs and WASP-70Ab transits the evolved primary of a G4+K3 binary},
  journal = {Monthly Notices of the Royal Astronomical Society},
  year    = {2014},
  volume  = {445},
  pages   = {1114--1129},
  doi     = {10.1093/mnras/stu1737}
}

@article{bonomo2017,
  author  = {Bonomo, Aldo S. and Desidera, Silvano and Benatti, Serena and Brosa, F. and Crespi, S. and Damasso, M.},
  title   = {The GAPS Programme with HARPS-N at TNG. XIV. Investigating giant planet migration history via improved eccentricity and mass determination for 231 transiting planets},
  journal = {Astronomy \& Astrophysics},
  year    = {2017},
  volume  = {602},
  pages   = {A107},
  doi     = {10.1051/0004-6361/201629882}
}

@article{ment2018,
  author  = {Ment, Kristo and Fischer, Debra A. and Bakos, Gaspar and Howard, Andrew H. and Isaacson, Howard},
  title   = {Radial Velocities from the N2K Project: Six New Cold Gas Giant Planets Orbiting HD 55696, HD 98736, HD 148164, HD 203473, and HD 211810},
  journal = {The Astronomical Journal},
  year    = {2018},
  volume  = {156},
  pages   = {213},
  doi={10.3847/1538-3881/aae1f5}
}

@article{sarkis2021,
  author  = {Sarkis, Paula and Mordasini, C. and Henning, Th. and Marleau, G. D. and Molliere, P.},
  title   = {Evidence of three mechanisms explaining the radius anomaly of hot Jupiters},
  journal = {Astronomy \& Astrophysics},
  year    = {2021},
  volume  = {645},
  pages   = {A79},
  doi     = {10.1051/0004-6361/202038361}
}

@article{bodenheimer2001,
  author  = {Bodenheimer, Peter and Lin, Douglas N. C. and Mardling, Rosemary A.},
  title   = {On the Tidal Inflation of Short-Period Extrasolar Planets},
  journal = {The Astrophysical Journal},
  year    = {2001},
  volume  = {548},
  pages   = {466--472},
  doi     = {10.1086/318667}
}

@article{batyginstevenson2010,
  author  = {Batygin, Konstantin and Stevenson, David J.},
  title   = {Inflating Hot Jupiters with Ohmic Dissipation},
  journal = {The Astrophysical Journal Letters},
  year    = {2010},
  volume  = {714},
  pages   = {L238--L243},
  doi     = {10.1088/2041-8205/714/2/L238}
}

@article{laughlin2011,
  author  = {Laughlin, Gregory and Crismani, Matteo and Adams, Fred C.},
  title   = {On the Anomalous Radii of the Transiting Extrasolar Planets},
  journal = {The Astrophysical Journal Letters},
  year    = {2011},
  volume  = {729},
  pages   = {L7},
  doi     = {10.1088/2041-8205/729/1/L7}
}

@article{thorngren2018,
  author  = {Thorngren, Daniel P. and Fortney, Jonathan J.},
  title   = {Bayesian Analysis of Hot-Jupiter Radius Anomalies: Evidence for Ohmic Dissipation?},
  journal = {The Astronomical Journal},
  year    = {2018},
  volume  = {155},
  pages   = {214},
  doi     = {10.3847/1538-3881/aaba13}
}

@article{huang2026,
  author  = {Huang, Helong and Min, Michiel and Ormel, Chris W. and Dyrek, Achrene and Crouzet, Nicolas},
  title   = {A Cloudy Fit to the Atmosphere of WASP-107 b},
  journal = {Astronomy \& Astrophysics},
  year    = {2026},
  volume  = {708},
  pages   = {L7},
  doi     = {10.1051/0004-6361/202558447},
}

@article{carleo2020,
  author  = {Carleo, Ilaria and Gandolfi, Davide and Barrag{\'a}n, Oscar and Livingston, John H. and Persson, Carina M. and others},
  title   = {The Multiplanet System TOI-421},
  journal = {The Astronomical Journal},
  year    = {2020},
  volume  = {160},
  pages   = {114},
  doi     = {10.3847/1538-3881/aba124}
}

@article{lindal1983,
  author  = {Lindal, Gunnar F. and Wood, G. E. and Hotz, H. B. and Sweetnam, D. N. and Eshleman, V. R. and Tyler, G. L.},
  title   = {The atmosphere of Titan: An analysis of the Voyager 1 radio occultation measurements},
  journal = {Icarus},
  year    = {1983},
  volume  = {53},
  pages   = {348--363},
  doi     = {10.1016/0019-1035(83)90155-0}
}

@article{lavvas2008,
  author  = {Lavvas, Panayotis P. and Coustenis, Athena and Vardavas, Ilias M.},
  title   = {Coupling photochemistry with haze formation in Titan's atmosphere, Part I: Model description},
  journal = {Planetary and Space Science},
  year    = {2008},
  volume  = {56},
  pages   = {27--66},
  doi     = {10.1016/j.pss.2007.05.026}
}

@article{tomasko2008,
  author  = {Tomasko, Martin G. and Doose, L. and Engel, S. and Dafoe, L. E. and West, R. and Lemmon, M. and others},
  title   = {A model of Titan's aerosols based on measurements made inside the atmosphere},
  journal = {Planetary and Space Science},
  year    = {2008},
  volume  = {56},
  pages   = {669--707},
  doi     = {10.1016/j.pss.2007.11.019}
}

@article{gladstone2016,
  author  = {Gladstone, G. Randall and Stern, S. Alan and Ennico, Kimberly and others},
  title   = {The atmosphere of Pluto as observed by New Horizons},
  journal = {Science},
  year    = {2016},
  volume  = {351},
  pages   = {aad8866},
  doi     = {10.1126/science.aad8866}
}

@article{gao2017,
  author  = {Gao, Peter and Fan, Siteng and Wong, Michael L. and Liang, Mao-Chang and Shia, Run-Lie and Kammer, Joshua A. and Yung, Yuk L. and others},
  title   = {Constraints on the microphysics of Pluto's photochemical haze from New Horizons observations},
  journal = {Icarus},
  year    = {2017},
  volume  = {287},
  pages   = {116--123},
  doi     = {10.1016/j.icarus.2016.09.030}
}

@article{arney2017,
  author  = {Arney, Giada N. and Meadows, Victoria S. and Domagal-Goldman, Shawn D. and Deming, Drake and Robinson, Tyler D. and Guadalupe, Tovar and others},
  title   = {Pale Orange Dots: The Impact of Organic Haze on the Habitability and Detectability of Archean Earth-like Exoplanets},
  journal = {The Astrophysical Journal},
  year    = {2017},
  volume  = {836},
  pages   = {49},
  doi     = {10.3847/1538-4357/836/1/49}
}

@article{adams2019,
  author  = {Adams, Danica and Gao, Peter and de Pater, Imke and Morley, Caroline V.},
  title   = {Aggregate Hazes in Exoplanet Atmospheres},
  journal = {The Astrophysical Journal},
  year    = {2019},
  volume  = {874},
  pages   = {61},
  doi     = {10.3847/1538-4357/ab074c}
}

@article{adams2022,
  author  = {Adams, Danica J. and Kataria, Tiffany and Batalha, Natasha E. and Gao, Peter and Knutson, Heather A.},
  title   = {Nightside Clouds and Disequilibrium Chemistry on the Hot Jupiter Kepler-7b},
  journal = {The Astrophysical Journal},
  year    = {2022},
  volume  = {926},
  pages   = {157},
  doi     = {10.3847/1538-4357/ac3d32}
}

@article{madhusudhan2012,
  author  = {Madhusudhan, Nikku},
  title   = {C/O Ratio as a Dimension for Characterizing Exoplanetary Atmospheres},
  journal = {The Astrophysical Journal},
  year    = {2012},
  volume  = {758},
  pages   = {36},
  doi     = {10.1088/0004-637X/758/1/36}
}

@article{fleury2020,
  author  = {Fleury, Benjamin and Gudipati, Murthy S. and Henderson, Bryana L. and Swain, Mark},
  title   = {Photochemistry in Hot H$_2$-dominated Exoplanet Atmospheres: Influence of the C/O Ratio},
  journal = {The Astrophysical Journal},
  year    = {2020},
  volume  = {899},
  pages   = {147},
  doi={10.3847/1538-4357/aaf79f}
}

@article{krasnopolsky1999,
  author  = {Krasnopolsky, Vladimir A. and Cruikshank, Dale P.},
  title   = {Photochemistry of Pluto's atmosphere and ionosphere near perihelion},
  journal = {Journal of Geophysical Research: Planets},
  year    = {1999},
  volume  = {104},
  pages   = {21979--21996},
  doi     = {10.1029/1999JE001038}
}

@article{wong2015,
  author  = {Wong, Michael L. and Yung, Yuk L. and Gladstone, G. Randall},
  title   = {Pluto's implications for a snowball Titan},
  journal = {Icarus},
  year    = {2015},
  volume  = {246},
  pages   = {192--196},
  doi={10.1016/j.icarus.2014.05.019}
}

@article{willacy2022,
  author  = {Willacy, Karen and Chen, Sihe and Adams, Danica and Yung, Yuk L.},
  title   = {Vertical Distribution of Cyclic Hydrocarbons and Nitriles in Titan's Atmosphere},
  journal = {The Astrophysical Journal},
  year    = {2022},
  volume={933},
  doi={10.3847/1538-4357/ac6b9d}
}

@article{rimmer2019,
  author  = {Rimmer, Paul B. and Rugheimer, Sarah},
  title   = {Hydrogen cyanide in nitrogen-rich atmospheres of rocky exoplanets},
  journal = {Icarus},
  year    = {2019},
  volume  = {329},
  pages   = {124--131},
  doi     = {10.1016/j.icarus.2019.02.020}
}

@article{friss2026,
  author  = {Friss, Gergely and Palmer, Paul I. and Braam, Marrick and Rice, Ken},
  title   = {Atmospheric Supply of Hydrogen Cyanide Is Not the Rate-limiting Step for Prebiotic Chemistry across Rocky Exoplanets},
  journal = {The Astrophysical Journal},
  year    = {2026},
  doi     = {10.3847/1538-4357/ae505d},
  volume={1001}
}

@article{kawashima2018,
  author  = {Kawashima, Yui and Ikoma, Masahiro},
  title   = {Theoretical Transmission Spectra of Exoplanet Atmospheres with Hydrocarbon Haze: Effect of Creation, Growth, and Settling of Haze Particles},
  journal = {The Astrophysical Journal},
  year    = {2018},
  volume  = {853},
  pages   = {7},
  doi     = {10.3847/1538-4357/aaa0c5}
}

@article{hobbs2019,
  author  = {Hobbs, Richard and others},
  title   = {A chemical kinetics code for modelling exoplanetary atmospheres},
  journal = {Monthly Notices of the Royal Astronomical Society},
  year    = {2019},
  volume  = {487},
  pages   = {2242--2261},
  doi={10.1093/mnras/stz1333}
}

@article{pearce2020,
  author  = {Pearce, Ben K. D. and Molaverdikhani, Karan and Pudritz, Ralph and Henning, Thomas and Hebrard, Eric},
  title   = {HCN production in Titan's atmosphere: Coupling quantum chemistry and disequilibrium atmospheric modeling},
  journal = {The Astrophysical Journal},
  year    = {2020},
  volume  = {901},
  pages   = {110},
  doi={10.3847/1538-4357/abae5c}
}

@article{ugelow2024,
  author  = {Ugelow, Melissa S. and Wieman, Scott T. and Schwarz, Madeline, C. R. and Da Poian, Victoria and Stern, Jennifer C. and Trainer, Melissa G. },
  title   = {Laboratory Studies on the Influence of Hydrogen on Titan-like Photochemistry},
  journal = {ACS Earth and Space Chemistry},
  year    = {2024},
  volume={8},
  doi={10.1021/acsearthspacechem.4c00102}
}

@article{veillet2024,
  author  = {Veillet, Rom{\'e}o and Venot, Olivia and Sirjean, Baptiste and Bounaceur, Roda and Glaude, Pierre-Alexandre and Al-Refaie, Ahmed and H{\'e}brard, Eric},
  title   = {An extensively validated C/H/O/N chemical network for hot exoplanet disequilibrium chemistry},
  journal = {Astronomy \& Astrophysics},
  year    = {2024},
  volume  = {682},
  pages   = {A52},
  doi     = {10.1051/0004-6361/202346680},
}

@article{veillet2026,
  author  = {Veillet, Rom{\'e}o and Venot, O. and Sirjean, B. and Destro, F. C. and Fournet, R. and others},
  title   = {Development of a C/H/O/N/S chemical network: Experimental benchmark, application to exoplanets, and identification of key C/S coupling pathways},
  journal = {Astronomy \& Astrophysics},
  year    = {2026},
  volume={706},
  doi     = {10.1051/0004-6361/202555595}
}

@article{moses_disequilibrium_2011,
	title = {{DISEQUILIBRIUM} {CARBON}, {OXYGEN}, {AND} {NITROGEN} {CHEMISTRY} {IN} {THE} {ATMOSPHERES} {OF} {HD} 189733b {AND} {HD} 209458b},
	volume = {737},
	issn = {0004-637X, 1538-4357},
	url = {https://iopscience.iop.org/article/10.1088/0004-637X/737/1/15},
	doi = {10.1088/0004-637X/737/1/15},
	language = {en},
	number = {1},
	urldate = {2020-07-10},
	journal = {The Astrophysical Journal},
	author = {Moses, Julianne I. and Visscher, C. and Fortney, J. J. and Showman, A. P. and Lewis, N. K. and Griffith, C. A. and Klippenstein, S. J. and Shabram, M. and Friedson, A. J. and Marley, M. S. and Freedman, R. S.},
	month = aug,
	year = {2011},
	pages = {15},
}

@article{moses_chemical_2013,
	title = {{CHEMICAL} {CONSEQUENCES} {OF} {THE} {C}/{O} {RATIO} {ON} {HOT} {JUPITERS}: {EXAMPLES} {FROM} {WASP}-12b, {CoRoT}-2b, {XO}-1b, {AND} {HD} 189733b},
	volume = {763},
	issn = {0004-637X, 1538-4357},
	shorttitle = {{CHEMICAL} {CONSEQUENCES} {OF} {THE} {C}/{O} {RATIO} {ON} {HOT} {JUPITERS}},
	url = {https://iopscience.iop.org/article/10.1088/0004-637X/763/1/25},
	doi = {10.1088/0004-637X/763/1/25},
	language = {en},
	number = {1},
	urldate = {2020-07-17},
	journal = {The Astrophysical Journal},
	author = {Moses, J. I. and Madhusudhan, N. and Visscher, C. and Freedman, R. S.},
	month = jan,
	year = {2013},
	pages = {25},
}

@article{pelletier_where_2021,
	title = {Where {Is} the {Water}? {Jupiter}-like {C}/{H} {Ratio} but {Strong} {H} $_{\textrm{2}}$ {O} {Depletion} {Found} on τ {Boötis} b {Using} {SPIRou}},
	volume = {162},
	issn = {0004-6256, 1538-3881},
	shorttitle = {Where {Is} the {Water}?},
	url = {https://iopscience.iop.org/article/10.3847/1538-3881/ac0428},
	doi = {10.3847/1538-3881/ac0428},
	language = {en},
	number = {2},
	urldate = {2021-09-15},
	journal = {The Astronomical Journal},
	author = {Pelletier, Stefan and Benneke, Björn and Darveau-Bernier, Antoine and Boucher, Anne and Cook, Neil J. and Piaulet, Caroline and Coulombe, Louis-Philippe and Artigau, Étienne and Lafrenière, David and Delisle, Simon and Allart, Romain and Doyon, René and Donati, Jean-François and Fouqué, Pascal and Moutou, Claire and Cadieux, Charles and Delfosse, Xavier and Hébrard, Guillaume and Martins, Jorge H. C. and Martioli, Eder and Vandal, Thomas},
	month = aug,
	year = {2021},
	pages = {73},
}

@article{benneke_atmospheric_2012,
	title = {{ATMOSPHERIC} {RETRIEVAL} {FOR} {SUPER}-{EARTHS}: {UNIQUELY} {CONSTRAINING} {THE} {ATMOSPHERIC} {COMPOSITION} {WITH} {TRANSMISSION} {SPECTROSCOPY}},
	volume = {753},
	issn = {0004-637X, 1538-4357},
	shorttitle = {{ATMOSPHERIC} {RETRIEVAL} {FOR} {SUPER}-{EARTHS}},
	url = {https://iopscience.iop.org/article/10.1088/0004-637X/753/2/100},
	doi = {10.1088/0004-637X/753/2/100},
	language = {en},
	number = {2},
	urldate = {2021-09-15},
	journal = {The Astrophysical Journal},
	author = {Benneke, Bjoern and Seager, Sara},
	month = jul,
	year = {2012},
	pages = {100},
}

@article{toon_rapid_1989,
	title = {Rapid calculation of radiative heating rates and photodissociation rates in inhomogeneous multiple scattering atmospheres},
	volume = {94},
	issn = {0148-0227},
	url = {http://doi.wiley.com/10.1029/JD094iD13p16287},
	doi = {10.1029/JD094iD13p16287},
	language = {en},
	number = {D13},
	urldate = {2021-09-15},
	journal = {Journal of Geophysical Research},
	author = {Toon, Owen B. and McKay, C. P. and Ackerman, T. P. and Santhanam, K.},
	year = {1989},
	pages = {16287},
}

@article{bonfils_hot_2012,
	title = {A hot {Uranus} transiting the nearby {M} dwarf {GJ} 3470: {Detected} with {HARPS} velocimetry. {Captured} in transit with {TRAPPIST} photometry⋆⋆⋆},
	volume = {546},
	issn = {0004-6361, 1432-0746},
	shorttitle = {A hot {Uranus} transiting the nearby {M} dwarf {GJ} 3470},
	url = {http://www.aanda.org/10.1051/0004-6361/201219623},
	doi = {10.1051/0004-6361/201219623},
	language = {en},
	urldate = {2021-09-20},
	journal = {Astronomy \& Astrophysics},
	author = {Bonfils, X. and Gillon, M. and Udry, S. and Armstrong, D. and Bouchy, F. and Delfosse, X. and Forveille, T. and Fumel, A. and Jehin, E. and Lendl, M. and Lovis, C. and Mayor, M. and McCormac, J. and Neves, V. and Pepe, F. and Perrier, C. and Pollaco, D. and Queloz, D. and Santos, N. C.},
	month = oct,
	year = {2012},
	pages = {A27},
}

@article{biddle_warm_2014,
	title = {Warm ice giant {GJ} 3470b - {II}. {Revised} planetary and stellar parameters from optical to near-infrared transit photometry},
	volume = {443},
	issn = {0035-8711, 1365-2966},
	url = {https://academic.oup.com/mnras/article-lookup/doi/10.1093/mnras/stu1199},
	doi = {10.1093/mnras/stu1199},
	language = {en},
	number = {2},
	urldate = {2021-09-20},
	journal = {Monthly Notices of the Royal Astronomical Society},
	author = {Biddle, L. I. and Pearson, K. A. and Crossfield, I. J. M. and Fulton, B. J. and Ciceri, S. and Eastman, J. and Barman, T. and Mann, A. W. and Henry, G. W. and Howard, A. W. and Williamson, M. H. and Sinukoff, E. and Dragomir, D. and Vican, L. and Mancini, L. and Southworth, J. and Greenberg, A. and Turner, J. D. and Thompson, R. and Taylor, B. W. and Levine, S. E. and Webber, M. W.},
	month = jul,
	year = {2014},
	pages = {1810--1820},
}

@article{stevenson_possible_2010,
	title = {Possible thermochemical disequilibrium in the atmosphere of the exoplanet {GJ} 436b},
	volume = {464},
	language = {en},
	author = {Stevenson, Kevin B},
	year = {2010},
	pages = {4},
}

@article{maas_lower-than-expected_2022,
	title = {Lower-than-expected flare temperatures for {TRAPPIST}-1},
	volume = {668},
	issn = {0004-6361, 1432-0746},
	url = {https://www.aanda.org/10.1051/0004-6361/202243869},
	doi = {10.1051/0004-6361/202243869},
	language = {en},
	urldate = {2023-01-19},
	journal = {Astronomy \& Astrophysics},
	author = {Maas, A. J. and Ilin, E. and Oshagh, M. and Pallé, E. and Parviainen, H. and Molaverdikhani, K. and Quirrenbach, A. and Esparza-Borges, E. and Murgas, F. and Béjar, V. J. S. and Narita, N. and Fukui, A. and Lin, C.-L. and Mori, M. and Klagyivik, P.},
	month = dec,
	year = {2022},
	pages = {A111},
}

@article{gao_aerosol_2020,
	title = {Aerosol composition of hot giant exoplanets dominated by silicates and hydrocarbon hazes},
	volume = {4},
	copyright = {2020 The Author(s), under exclusive licence to Springer Nature Limited},
	issn = {2397-3366},
	url = {https://www.nature.com/articles/s41550-020-1114-3},
	doi = {10.1038/s41550-020-1114-3},
	language = {en},
	number = {10},
	urldate = {2023-01-27},
	journal = {Nature Astronomy},
	publisher = {Nature Publishing Group},
	author = {Gao, Peter and Thorngren, Daniel P. and Lee, Elspeth K. H. and Fortney, Jonathan J. and Morley, Caroline V. and Wakeford, Hannah R. and Powell, Diana K. and Stevenson, Kevin B. and Zhang, Xi},
	month = oct,
	year = {2020},
	note = {Number: 10},
	pages = {951--956},
}

@inproceedings{skilling_nested_2004,
	address = {Garching (Germany)},
	title = {Nested {Sampling}},
	volume = {735},
	issn = {0094243X},
	url = {http://aip.scitation.org/doi/abs/10.1063/1.1835238},
	doi = {10.1063/1.1835238},
	language = {en},
	urldate = {2023-04-19},
	booktitle = {{AIP} {Conference} {Proceedings}},
	publisher = {AIP},
	author = {Skilling, John},
	year = {2004},
	note = {tex.editor: \{\}},
	pages = {395--405},
}

@article{ahrer_early_2023,
	title = {Early {Release} {Science} of the exoplanet {WASP}-39b with {JWST} {NIRCam}},
	volume = {614},
	copyright = {2023 The Author(s)},
	issn = {1476-4687},
	url = {https://www.nature.com/articles/s41586-022-05590-4},
	doi = {10.1038/s41586-022-05590-4},
	language = {en},
	number = {7949},
	urldate = {2023-04-20},
	journal = {Nature},
	publisher = {Nature Publishing Group},
	author = {Ahrer, Eva-Maria and Stevenson, Kevin B. and Mansfield, Megan and Moran, Sarah E. and Brande, Jonathan and Morello, Giuseppe and Murray, Catriona A. and Nikolov, Nikolay K. and Petit dit de la Roche, Dominique J. M. and Schlawin, Everett and Wheatley, Peter J. and Zieba, Sebastian and Batalha, Natasha E. and Damiano, Mario and Goyal, Jayesh M. and Lendl, Monika and Lothringer, Joshua D. and Mukherjee, Sagnick and Ohno, Kazumasa and Batalha, Natalie M. and Battley, Matthew P. and Bean, Jacob L. and Beatty, Thomas G. and Benneke, Björn and Berta-Thompson, Zachory K. and Carter, Aarynn L. and Cubillos, Patricio E. and Daylan, Tansu and Espinoza, Néstor and Gao, Peter and Gibson, Neale P. and Gill, Samuel and Harrington, Joseph and Hu, Renyu and Kreidberg, Laura and Lewis, Nikole K. and Line, Michael R. and López-Morales, Mercedes and Parmentier, Vivien and Powell, Diana K. and Sing, David K. and Tsai, Shang-Min and Wakeford, Hannah R. and Welbanks, Luis and Alam, Munazza K. and Alderson, Lili and Allen, Natalie H. and Anderson, David R. and Barstow, Joanna K. and Bayliss, Daniel and Bell, Taylor J. and Blecic, Jasmina and Bryant, Edward M. and Burleigh, Matthew R. and Carone, Ludmila and Casewell, S. L. and Changeat, Quentin and Chubb, Katy L. and Crossfield, Ian J. M. and Crouzet, Nicolas and Decin, Leen and Désert, Jean-Michel and Feinstein, Adina D. and Flagg, Laura and Fortney, Jonathan J. and Gizis, John E. and Heng, Kevin and Iro, Nicolas and Kempton, Eliza M.-R. and Kendrew, Sarah and Kirk, James and Knutson, Heather A. and Komacek, Thaddeus D. and Lagage, Pierre-Olivier and Leconte, Jérémy and Lustig-Yaeger, Jacob and MacDonald, Ryan J. and Mancini, Luigi and May, E. M. and Mayne, N. J. and Miguel, Yamila and Mikal-Evans, Thomas and Molaverdikhani, Karan and Palle, Enric and Piaulet, Caroline and Rackham, Benjamin V. and Redfield, Seth and Rogers, Laura K. and Roy, Pierre-Alexis and Rustamkulov, Zafar and Shkolnik, Evgenya L. and Sotzen, Kristin S. and Taylor, Jake and Tremblin, P. and Tucker, Gregory S. and Turner, Jake D. and de Val-Borro, Miguel and Venot, Olivia and Zhang, Xi},
	month = feb,
	year = {2023},
	note = {Number: 7949},
	pages = {653--658},
}

@article{charbonneau_detection_2000,
	title = {Detection of {Planetary} {Transits} {Across} a {Sun}-like {Star}},
	volume = {529},
	issn = {0004637X},
	url = {https://iopscience.iop.org/article/10.1086/312457},
	doi = {10.1086/312457},
	language = {en},
	number = {1},
	urldate = {2023-04-24},
	journal = {The Astrophysical Journal},
	author = {Charbonneau, David and Brown, Timothy M. and Latham, David W. and Mayor, Michel},
	month = jan,
	year = {2000},
	pages = {L45--L48},
}

@article{benneke_water_2019,
	title = {Water {Vapor} and {Clouds} on the {Habitable}-zone {Sub}-{Neptune} {Exoplanet} {K2}-18b},
	volume = {887},
	issn = {2041-8205, 2041-8213},
	url = {https://iopscience.iop.org/article/10.3847/2041-8213/ab59dc},
	doi = {10.3847/2041-8213/ab59dc},
	language = {en},
	number = {1},
	urldate = {2023-05-19},
	journal = {The Astrophysical Journal Letters},
	author = {Benneke, Björn and Wong, Ian and Piaulet, Caroline and Knutson, Heather A. and Lothringer, Joshua and Morley, Caroline V. and Crossfield, Ian J. M. and Gao, Peter and Greene, Thomas P. and Dressing, Courtney and Dragomir, Diana and Howard, Andrew W. and McCullough, Peter R. and Kempton, Eliza M.-R. and Fortney, Jonathan J. and Fraine, Jonathan},
	month = dec,
	year = {2019},
	pages = {L14},
}

@article{swain_presence_2008,
	title = {The presence of methane in the atmosphere of an extrasolar planet},
	volume = {452},
	copyright = {2008 Springer Nature Limited},
	issn = {1476-4687},
	url = {https://www.nature.com/articles/nature06823},
	doi = {10.1038/nature06823},
	language = {en},
	number = {7185},
	urldate = {2023-06-21},
	journal = {Nature},
	publisher = {Nature Publishing Group},
	author = {Swain, Mark R. and Vasisht, Gautam and Tinetti, Giovanna},
	month = mar,
	year = {2008},
	note = {Number: 7185},
	pages = {329--331},
}

@article{roy_water_2023,
	title = {Water {Absorption} in the {Transmission} {Spectrum} of the {Water} {World} {Candidate} {GJ} 9827 d},
	volume = {954},
	issn = {2041-8205, 2041-8213},
	url = {https://iopscience.iop.org/article/10.3847/2041-8213/acebf0},
	doi = {10.3847/2041-8213/acebf0},
	language = {en},
	number = {2},
	urldate = {2023-09-19},
	journal = {The Astrophysical Journal Letters},
	author = {Roy, Pierre-Alexis and Benneke, Björn and Piaulet, Caroline and Gully-Santiago, Michael A. and Crossfield, Ian J. M. and Morley, Caroline V. and Kreidberg, Laura and Mikal-Evans, Thomas and Brande, Jonathan and Delisle, Simon and Greene, Thomas P. and Hardegree-Ullman, Kevin K. and Barman, Travis and Christiansen, Jessie L. and Dragomir, Diana and Fortney, Jonathan J. and Howard, Andrew W. and Kosiarek, Molly R. and Lothringer, Joshua D.},
	month = sep,
	year = {2023},
	pages = {L52},
}

@article{benneke_sub-neptune_2019,
	title = {A sub-{Neptune} exoplanet with a low-metallicity methane-depleted atmosphere and {Mie}-scattering clouds},
	volume = {3},
	copyright = {2019 The Author(s), under exclusive licence to Springer Nature Limited},
	issn = {2397-3366},
	url = {https://www.nature.com/articles/s41550-019-0800-5},
	doi = {10.1038/s41550-019-0800-5},
	language = {en},
	number = {9},
	urldate = {2023-10-31},
	journal = {Nature Astronomy},
	publisher = {Nature Publishing Group},
	author = {Benneke, Björn and Knutson, Heather A. and Lothringer, Joshua and Crossfield, Ian J. M. and Moses, Julianne I. and Morley, Caroline and Kreidberg, Laura and Fulton, Benjamin J. and Dragomir, Diana and Howard, Andrew W. and Wong, Ian and Désert, Jean-Michel and McCullough, Peter R. and Kempton, Eliza M.-R. and Fortney, Jonathan and Gilliland, Ronald and Deming, Drake and Kammer, Joshua},
	month = sep,
	year = {2019},
	note = {Number: 9},
	pages = {813--821},
}

@article{hu_photochemistry_2021,
	title = {Photochemistry and {Spectral} {Characterization} of {Temperate} and {Gas}-rich {Exoplanets}},
	volume = {921},
	issn = {0004-637X, 1538-4357},
	url = {https://iopscience.iop.org/article/10.3847/1538-4357/ac1789},
	doi = {10.3847/1538-4357/ac1789},
	language = {en},
	number = {1},
	urldate = {2023-11-16},
	journal = {The Astrophysical Journal},
	author = {Hu, Renyu},
	month = nov,
	year = {2021},
	pages = {27},
}

@article{madhusudhan_carbon-bearing_2023,
	title = {Carbon-bearing {Molecules} in a {Possible} {Hycean} {Atmosphere}},
	volume = {956},
	issn = {2041-8205, 2041-8213},
	url = {https://iopscience.iop.org/article/10.3847/2041-8213/acf577},
	doi = {10.3847/2041-8213/acf577},
	language = {en},
	number = {1},
	urldate = {2024-01-08},
	journal = {The Astrophysical Journal Letters},
	author = {Madhusudhan, Nikku and Sarkar, Subhajit and Constantinou, Savvas and Holmberg, Måns and Piette, Anjali A. A. and Moses, Julianne I.},
	month = oct,
	year = {2023},
	pages = {L13},
}

@article{howard_characterizing_2023,
	title = {Characterizing the {Near}-infrared {Spectra} of {Flares} from {TRAPPIST}-1 during {JWST} {Transit} {Spectroscopy} {Observations}},
	volume = {959},
	issn = {0004-637X, 1538-4357},
	url = {https://iopscience.iop.org/article/10.3847/1538-4357/acfe75},
	doi = {10.3847/1538-4357/acfe75},
	language = {en},
	number = {1},
	urldate = {2024-01-15},
	journal = {The Astrophysical Journal},
	author = {Howard, Ward S. and Kowalski, Adam F. and Flagg, Laura and MacGregor, Meredith A. and Lim, Olivia and Radica, Michael and Piaulet, Caroline and Roy, Pierre-Alexis and Lafrenière, David and Benneke, Björn and Brown, Alexander and Espinoza, Néstor and Doyon, René and Coulombe, Louis-Philippe and Johnstone, Doug and Cowan, Nicolas B. and Jayawardhana, Ray and Turner, Jake D. and Dang, Lisa},
	month = dec,
	year = {2023},
	pages = {64},
}

@article{gunther_super-earth_2019,
	title = {A super-{Earth} and two sub-{Neptunes} transiting the nearby and quiet {M} dwarf {TOI}-270},
	volume = {3},
	copyright = {2019 The Author(s), under exclusive licence to Springer Nature Limited},
	issn = {2397-3366},
	url = {https://www.nature.com/articles/s41550-019-0845-5},
	doi = {10.1038/s41550-019-0845-5},
	language = {en},
	number = {12},
	urldate = {2024-02-23},
	journal = {Nature Astronomy},
	publisher = {Nature Publishing Group},
	author = {Günther, Maximilian N. and Pozuelos, Francisco J. and Dittmann, Jason A. and Dragomir, Diana and Kane, Stephen R. and Daylan, Tansu and Feinstein, Adina D. and Huang, Chelsea X. and Morton, Timothy D. and Bonfanti, Andrea and Bouma, L. G. and Burt, Jennifer and Collins, Karen A. and Lissauer, Jack J. and Matthews, Elisabeth and Montet, Benjamin T. and Vanderburg, Andrew and Wang, Songhu and Winters, Jennifer G. and Ricker, George R. and Vanderspek, Roland K. and Latham, David W. and Seager, Sara and Winn, Joshua N. and Jenkins, Jon M. and Armstrong, James D. and Barkaoui, Khalid and Batalha, Natalie and Bean, Jacob L. and Caldwell, Douglas A. and Ciardi, David R. and Collins, Kevin I. and Crossfield, Ian and Fausnaugh, Michael and Furesz, Gabor and Gan, Tianjun and Gillon, Michaël and Guerrero, Natalia and Horne, Keith and Howell, Steve B. and Ireland, Michael and Isopi, Giovanni and Jehin, Emmanuël and Kielkopf, John F. and Lepine, Sebastien and Mallia, Franco and Matson, Rachel A. and Myers, Gordon and Palle, Enric and Quinn, Samuel N. and Relles, Howard M. and Rojas-Ayala, Bárbara and Schlieder, Joshua and Sefako, Ramotholo and Shporer, Avi and Suárez, Juan C. and Tan, Thiam-Guan and Ting, Eric B. and Twicken, Joseph D. and Waite, Ian A.},
	month = dec,
	year = {2019},
	note = {Number: 12},
	pages = {1099--1108},
}

@article{tsai_comparative_2021,
	title = {A {Comparative} {Study} of {Atmospheric} {Chemistry} with {VULCAN}},
	volume = {923},
	issn = {0004-637X, 1538-4357},
	url = {https://iopscience.iop.org/article/10.3847/1538-4357/ac29bc},
	doi = {10.3847/1538-4357/ac29bc},
	language = {en},
	number = {2},
	urldate = {2024-04-16},
	journal = {The Astrophysical Journal},
	author = {Tsai, Shang-Min and Malik, Matej and Kitzmann, Daniel and Lyons, James R. and Fateev, Alexander and Lee, Elspeth and Heng, Kevin},
	month = dec,
	year = {2021},
	pages = {264},
}

@article{welbanks_high_2024,
	title = {A high internal heat flux and large core in a warm {Neptune} exoplanet},
	volume = {630},
	copyright = {2024 The Author(s), under exclusive licence to Springer Nature Limited},
	issn = {1476-4687},
	url = {https://www.nature.com/articles/s41586-024-07514-w},
	doi = {10.1038/s41586-024-07514-w},
	language = {en},
	number = {8018},
	urldate = {2024-07-17},
	journal = {Nature},
	publisher = {Nature Publishing Group},
	author = {Welbanks, Luis and Bell, Taylor J. and Beatty, Thomas G. and Line, Michael R. and Ohno, Kazumasa and Fortney, Jonathan J. and Schlawin, Everett and Greene, Thomas P. and Rauscher, Emily and McGill, Peter and Murphy, Matthew and Parmentier, Vivien and Tang, Yao and Edelman, Isaac and Mukherjee, Sagnick and Wiser, Lindsey S. and Lagage, Pierre-Olivier and Dyrek, Achrène and Arnold, Kenneth E.},
	month = jun,
	year = {2024},
	pages = {836--840},
}

@misc{benneke_jwst_2024,
	title = {{JWST} {Reveals} {CH}\$\_4\$, {CO}\$\_2\$, and {H}\$\_2\${O} in a {Metal}-rich {Miscible} {Atmosphere} on a {Two}-{Earth}-{Radius} {Exoplanet}},
	url = {http://arxiv.org/abs/2403.03325},
	language = {en},
	urldate = {2024-08-06},
	publisher = {arXiv},
	author = {Benneke, Björn and Roy, Pierre-Alexis and Coulombe, Louis-Philippe and Radica, Michael and Piaulet, Caroline and Ahrer, Eva-Maria and Pierrehumbert, Raymond and Krissansen-Totton, Joshua and Schlichting, Hilke E. and Hu, Renyu and Yang, Jeehyun and Christie, Duncan and Thorngren, Daniel and Young, Edward D. and Pelletier, Stefan and Knutson, Heather A. and Miguel, Yamila and Evans-Soma, Thomas M. and Dorn, Caroline and Gagnebin, Anna and Fortney, Jonathan J. and Komacek, Thaddeus and MacDonald, Ryan and Raul, Eshan and Cloutier, Ryan and Acuna, Lorena and Lafrenière, David and Cadieux, Charles and Doyon, René and Welbanks, Luis and Allart, Romain},
	month = mar,
	year = {2024},
	note = {arXiv:2403.03325 [astro-ph]},
}

@article{bell_methane_2023,
	title = {Methane throughout the atmosphere of the warm exoplanet {WASP}-80b},
	volume = {623},
	copyright = {2023 The Author(s), under exclusive licence to Springer Nature Limited},
	issn = {1476-4687},
	url = {https://www.nature.com/articles/s41586-023-06687-0},
	doi = {10.1038/s41586-023-06687-0},
	language = {en},
	number = {7988},
	urldate = {2024-09-03},
	journal = {Nature},
	publisher = {Nature Publishing Group},
	author = {Bell, Taylor J. and Welbanks, Luis and Schlawin, Everett and Line, Michael R. and Fortney, Jonathan J. and Greene, Thomas P. and Ohno, Kazumasa and Parmentier, Vivien and Rauscher, Emily and Beatty, Thomas G. and Mukherjee, Sagnick and Wiser, Lindsey S. and Boyer, Martha L. and Rieke, Marcia J. and Stansberry, John A.},
	month = nov,
	year = {2023},
	pages = {709--712},
}

@article{fournier-tondreau_near-infrared_2024,
	title = {Near-infrared transmission spectroscopy of {HAT}-{P}-18 b with {NIRISS}: {Disentangling} planetary and stellar features in the era of \textit{{JWST}}},
	volume = {528},
	copyright = {https://creativecommons.org/licenses/by/4.0/},
	issn = {0035-8711, 1365-2966},
	shorttitle = {Near-infrared transmission spectroscopy of {HAT}-{P}-18 b with {NIRISS}},
	url = {https://academic.oup.com/mnras/article/528/2/3354/7468143},
	doi = {10.1093/mnras/stad3813},
	language = {en},
	number = {2},
	urldate = {2024-09-13},
	journal = {Monthly Notices of the Royal Astronomical Society},
	author = {Fournier-Tondreau, Marylou and MacDonald, Ryan J and Radica, Michael and Lafrenière, David and Welbanks, Luis and Piaulet, Caroline and Coulombe, Louis-Philippe and Allart, Romain and Morel, Kim and Artigau, Étienne and Albert, Loïc and Lim, Olivia and Doyon, René and Benneke, Björn and Rowe, Jason F and Darveau-Bernier, Antoine and Cowan, Nicolas B and Lewis, Nikole K and Cook, Neil J and Flagg, Laura and Genest, Frédéric and Pelletier, Stefan and Johnstone, Doug and Dang, Lisa and Kaltenegger, Lisa and Taylor, Jake and Turner, Jake D},
	month = jan,
	year = {2024},
	pages = {3354--3377},
}

@article{gunther_stellar_2020,
	title = {Stellar {Flares} from the {First} {TESS} {Data} {Release}: {Exploring} a {New} {Sample} of {M} {Dwarfs}},
	volume = {159},
	issn = {0004-6256, 1538-3881},
	shorttitle = {Stellar {Flares} from the {First} {TESS} {Data} {Release}},
	url = {https://iopscience.iop.org/article/10.3847/1538-3881/ab5d3a},
	doi = {10.3847/1538-3881/ab5d3a},
	language = {en},
	number = {2},
	urldate = {2024-09-16},
	journal = {The Astronomical Journal},
	author = {Günther, Maximilian N. and Zhan, Zhuchang and Seager, Sara and Rimmer, Paul B. and Ranjan, Sukrit and Stassun, Keivan G. and Oelkers, Ryan J. and Daylan, Tansu and Newton, Elisabeth and Kristiansen, Martti H. and Olah, Katalin and Gillen, Edward and Rappaport, Saul and Ricker, George R. and Vanderspek, Roland K. and Latham, David W. and Winn, Joshua N. and Jenkins, Jon M. and Glidden, Ana and Fausnaugh, Michael and Levine, Alan M. and Dittmann, Jason A. and Quinn, Samuel N. and Krishnamurthy, Akshata and Ting, Eric B.},
	month = feb,
	year = {2020},
	pages = {60},
}

@article{beatty_sulfur_2024,
	title = {Sulfur {Dioxide} and {Other} {Molecular} {Species} in the {Atmosphere} of the {Sub}-{Neptune} {GJ} 3470 b},
	volume = {970},
	issn = {2041-8205, 2041-8213},
	url = {https://iopscience.iop.org/article/10.3847/2041-8213/ad55e9},
	doi = {10.3847/2041-8213/ad55e9},
	language = {en},
	number = {1},
	urldate = {2024-09-22},
	journal = {The Astrophysical Journal Letters},
	author = {Beatty, Thomas G. and Welbanks, Luis and Schlawin, Everett and Bell, Taylor J. and Line, Michael R. and Murphy, Matthew and Edelman, Isaac and Greene, Thomas P. and Fortney, Jonathan J. and Henry, Gregory W. and Mukherjee, Sagnick and Ohno, Kazumasa and Parmentier, Vivien and Rauscher, Emily and Wiser, Lindsey S. and Arnold, Kenneth E.},
	month = jul,
	year = {2024},
	pages = {L10},
}

@article{fu_hydrogen_2024,
	title = {Hydrogen sulfide and metal-enriched atmosphere for a {Jupiter}-mass exoplanet},
	volume = {632},
	copyright = {2024 The Author(s), under exclusive licence to Springer Nature Limited},
	issn = {1476-4687},
	url = {https://www.nature.com/articles/s41586-024-07760-y},
	doi = {10.1038/s41586-024-07760-y},
	language = {en},
	number = {8026},
	urldate = {2024-10-03},
	journal = {Nature},
	publisher = {Nature Publishing Group},
	author = {Fu, Guangwei and Welbanks, Luis and Deming, Drake and Inglis, Julie and Zhang, Michael and Lothringer, Joshua and Ih, Jegug and Moses, Julianne I. and Schlawin, Everett and Knutson, Heather A. and Henry, Gregory and Greene, Thomas and Sing, David K. and Savel, Arjun B. and Kempton, Eliza M.-R. and Louie, Dana R. and Line, Michael and Nixon, Matt},
	month = aug,
	year = {2024},
	pages = {752--756},
}

@article{piaulet-ghorayeb_jwstniriss_2024,
	title = {{JWST}/{NIRISS} {Reveals} the {Water}-rich “{Steam} {World}” {Atmosphere} of {GJ} 9827 d},
	volume = {974},
	issn = {2041-8205, 2041-8213},
	url = {https://iopscience.iop.org/article/10.3847/2041-8213/ad6f00},
	doi = {10.3847/2041-8213/ad6f00},
	language = {en},
	number = {1},
	urldate = {2024-10-16},
	journal = {The Astrophysical Journal Letters},
	author = {Piaulet-Ghorayeb, Caroline and Benneke, Björn and Radica, Michael and Raul, Eshan and Coulombe, Louis-Philippe and Ahrer, Eva-Maria and Kubyshkina, Daria and Howard, Ward S. and Krissansen-Totton, Joshua and MacDonald, Ryan J. and Roy, Pierre-Alexis and Louca, Amy and Christie, Duncan and Fournier-Tondreau, Marylou and Allart, Romain and Miguel, Yamila and Schlichting, Hilke E. and Welbanks, Luis and Cadieux, Charles and Dorn, Caroline and Evans-Soma, Thomas M. and Fortney, Jonathan J. and Pierrehumbert, Raymond and Lafrenière, David and Acuña, Lorena and Komacek, Thaddeus and Innes, Hamish and Beatty, Thomas G. and Cloutier, Ryan and Doyon, René and Gagnebin, Anna and Gapp, Cyril and Knutson, Heather A.},
	month = oct,
	year = {2024},
	pages = {L10},
}

@article{sing_warm_2024,
	title = {A warm {Neptune}’s methane reveals core mass and vigorous atmospheric mixing},
	volume = {630},
	copyright = {2024 The Author(s)},
	issn = {1476-4687},
	url = {https://www.nature.com/articles/s41586-024-07395-z},
	doi = {10.1038/s41586-024-07395-z},
	language = {en},
	number = {8018},
	urldate = {2024-10-22},
	journal = {Nature},
	publisher = {Nature Publishing Group},
	author = {Sing, David K. and Rustamkulov, Zafar and Thorngren, Daniel P. and Barstow, Joanna K. and Tremblin, Pascal and Alves de Oliveira, Catarina and Beck, Tracy L. and Birkmann, Stephan M. and Challener, Ryan C. and Crouzet, Nicolas and Espinoza, Néstor and Ferruit, Pierre and Giardino, Giovanna and Gressier, Amélie and Lee, Elspeth K. H. and Lewis, Nikole K. and Maiolino, Roberto and Manjavacas, Elena and Rauscher, Bernard J. and Sirianni, Marco and Valenti, Jeff A.},
	month = jun,
	year = {2024},
	pages = {831--835},
}

@misc{alam_jwst_2024,
	title = {{JWST} {COMPASS}: {The} first near- to mid-infrared transmission spectrum of the hot super-{Earth} {L} 168-9 b},
	shorttitle = {{JWST} {COMPASS}},
	url = {http://arxiv.org/abs/2411.03154},
	language = {en},
	urldate = {2024-11-06},
	publisher = {arXiv},
	author = {Alam, Munazza K. and Gao, Peter and Redai, Jea Adams and Wallack, Nicole L. and Wogan, Nicholas F. and Aguichine, Artyom and Dattilo, Anne and Alderson, Lili and Batalha, Natasha E. and Batalha, Natalie M. and Kirk, James and López-Morales, Mercedes and Meech, Annabella and Moran, Sarah E. and Teske, Johanna and Wakeford, Hannah R. and Wolfgang, Angie},
	month = nov,
	year = {2024},
	note = {arXiv:2411.03154 [astro-ph]},
}

@article{krenn_characterisation_2024,
	title = {Characterisation of the {TOI}-421 planetary system using {CHEOPS}, {TESS}, and archival radial velocity data},
	volume = {686},
	copyright = {https://creativecommons.org/licenses/by/4.0},
	issn = {0004-6361, 1432-0746},
	url = {https://www.aanda.org/10.1051/0004-6361/202348584},
	doi = {10.1051/0004-6361/202348584},
	language = {en},
	urldate = {2025-02-13},
	journal = {Astronomy \& Astrophysics},
	author = {Krenn, A. F. and Kubyshkina, D. and Fossati, L. and Egger, J. A. and Bonfanti, A. and Deline, A. and Ehrenreich, D. and Beck, M. and Benz, W. and Cabrera, J. and Wilson, T. G. and Leleu, A. and Sousa, S. G. and Adibekyan, V. and Correia, A. C. M. and Alibert, Y. and Delrez, L. and Lendl, M. and Patel, J. A. and Venturini, J. and Alonso, R. and Anglada, G. and Asquier, J. and Bárczy, T. and Barrado Navascues, D. and Barros, S. C. C. and Baumjohann, W. and Beck, T. and Billot, N. and Bonfils, X. and Borsato, L. and Brandeker, A. and Broeg, C. and Charnoz, S. and Collier Cameron, A. and Csizmadia, Sz. and Cubillos, P. E. and Davies, M. B. and Deleuil, M. and Demangeon, O. D. S. and Demory, B.-O. and Erikson, A. and Fortier, A. and Fridlund, M. and Gandolfi, D. and Gillon, M. and Güdel, M. and Günther, M. N. and Hasiba, J. and Heitzmann, A. and Helling, C. and Hoyer, S. and Isaak, K. G. and Kiss, L. L. and Lam, K. W. F. and Laskar, J. and Lecavelier Des Etangs, A. and Lovis, C. and Magrin, D. and Maxted, P. F. L. and Mordasini, C. and Nascimbeni, V. and Olofsson, G. and Ottensamer, R. and Pagano, I. and Pallé, E. and Peter, G. and Piotto, G. and Pollacco, D. and Queloz, D. and Ragazzoni, R. and Rando, N. and Rauer, H. and Ribas, I. and Rieder, M. and Santos, N. C. and Scandariato, G. and Ségransan, D. and Simon, A. E. and Smith, A. M. S. and Stalport, M. and Steller, M. and Szabó, Gy. M. and Thomas, N. and Udry, S. and Ulmer, B. and Van Grootel, V. and Villaver, E. and Viotto, V. and Walton, N. A. and Zingales, T.},
	month = jun,
	year = {2024},
	pages = {A301},
}

@article{holmberg_possible_2024,
	title = {Possible {Hycean} conditions in the sub-{Neptune} {TOI}-270 d},
	volume = {683},
	copyright = {© The Authors 2024},
	issn = {0004-6361, 1432-0746},
	url = {https://www.aanda.org/articles/aa/abs/2024/03/aa48238-23/aa48238-23.html},
	doi = {10.1051/0004-6361/202348238},
	language = {en},
	urldate = {2025-02-20},
	journal = {Astronomy \& Astrophysics},
	publisher = {EDP Sciences},
	author = {Holmberg, Måns and Madhusudhan, Nikku},
	month = mar,
	year = {2024},
	pages = {L2},
}

@article{benneke_how_2013,
	title = {{HOW} {TO} {DISTINGUISH} {BETWEEN} {CLOUDY} {MINI}-{NEPTUNES} {AND} {WATER}/{VOLATILE}-{DOMINATED} {SUPER}-{EARTHS}},
	volume = {778},
	issn = {0004-637X},
	url = {https://dx.doi.org/10.1088/0004-637X/778/2/153},
	doi = {10.1088/0004-637X/778/2/153},
	language = {en},
	number = {2},
	urldate = {2025-07-04},
	journal = {The Astrophysical Journal},
	publisher = {The American Astronomical Society},
	author = {Benneke, Björn and Seager, Sara},
	month = nov,
	year = {2013},
	pages = {153},
}

@misc{hu_water-rich_2025,
	title = {A water-rich interior in the temperate sub-{Neptune} {K2}-18 b revealed by {JWST}},
	url = {http://arxiv.org/abs/2507.12622},
	doi = {10.48550/arXiv.2507.12622},
	urldate = {2025-07-25},
	publisher = {arXiv},
	author = {Hu, Renyu and Bello-Arufe, Aaron and Tokadjian, Armen and Yang, Jeehyun and Damiano, Mario and Roy, Pierre-Alexis and Coulombe, Louis-Philippe and Madhusudhan, Nikku and Constantinou, Savvas and Benneke, Björn},
	month = jul,
	year = {2025},
	note = {arXiv:2507.12622 [astro-ph]
version: 1},
}

@article{wogan_jwst_2024,
	title = {{JWST} {Observations} of {K2}-18b {Can} {Be} {Explained} by a {Gas}-rich {Mini}-{Neptune} with {No} {Habitable} {Surface}},
	volume = {963},
	issn = {2041-8205},
	url = {https://doi.org/10.3847/2041-8213/ad2616},
	doi = {10.3847/2041-8213/ad2616},
	language = {en},
	number = {1},
	urldate = {2025-10-08},
	journal = {The Astrophysical Journal Letters},
	publisher = {The American Astronomical Society},
	author = {Wogan, Nicholas F. and Batalha, Natasha E. and Zahnle, Kevin J. and Krissansen-Totton, Joshua and Tsai, Shang-Min and Hu, Renyu},
	month = feb,
	year = {2024},
	pages = {L7},
}

@article{tsai_vulcan_2017,
	title = {{VULCAN}: {An} {Open}-source, {Validated} {Chemical} {Kinetics} {Python} {Code} for {Exoplanetary} {Atmospheres}},
	volume = {228},
	issn = {0067-0049},
	shorttitle = {{VULCAN}},
	url = {https://doi.org/10.3847/1538-4365/228/2/20},
	doi = {10.3847/1538-4365/228/2/20},
	language = {en},
	number = {2},
	urldate = {2025-10-08},
	journal = {The Astrophysical Journal Supplement Series},
	publisher = {The American Astronomical Society},
	author = {Tsai, Shang-Min and Lyons, James R. and Grosheintz, Luc and Rimmer, Paul B. and Kitzmann, Daniel and Heng, Kevin},
	month = feb,
	year = {2017},
	pages = {20},
}
\bibliographystyle{aasjournalv7}



\end{document}